\documentclass[11pt]{article}

\usepackage[top=1in,bottom=1in,left=1in,right=1in]{geometry}
\usepackage{natbib}
\usepackage[unicode=true,bookmarks=true,bookmarksnumbered=false,bookmarksopen=false,breaklinks=false,pdfborder={0 0 1},backref=false,colorlinks=true]{hyperref}
\hypersetup{citecolor=blue}
\usepackage{url}

\usepackage{amsmath}
\usepackage{mathrsfs}
\usepackage{amssymb}
\usepackage{amsthm}
\usepackage[normalem]{ulem}

\newtheorem{theorem}{Theorem}
\newtheorem{lemma}{Lemma}

\newtheorem{proposition}{Proposition}

\usepackage{float}
\usepackage{rotfloat}

\floatstyle{ruled}
\newfloat{algorithm}{tpb}{loa}
\providecommand{\algorithmname}{Algorithm}
\floatname{algorithm}{\protect\algorithmname}

\usepackage{algpseudocode,algorithm}
\algnewcommand\algorithmicinput{\textbf{Input}:}
\algnewcommand\algorithmicoutput{\textbf{Output}:}
\algnewcommand\INPUT{\item[\algorithmicinput]}
\algnewcommand\OUTPUT{\item[\algorithmicoutput]}

\usepackage{graphicx}
\usepackage[caption=false,labelformat=simple]{subfig}

\graphicspath{{/Users/yz125/Dropbox/MyFolder/Biostat-IU/Projects/cap/longitudinal/gamma_var/}}

\usepackage{array}
\newcolumntype{L}[1]{>{\raggedright\let\newline\\\arraybackslash\hspace{0pt}}m{#1}}
\newcolumntype{C}[1]{>{\centering\let\newline\\\arraybackslash\hspace{0pt}}m{#1}}
\newcolumntype{R}[1]{>{\raggedleft\let\newline\\\arraybackslash\hspace{0pt}}m{#1}}

\usepackage{rotating}
\usepackage{multirow}

\usepackage{tikz}
\usetikzlibrary{shapes,arrows,backgrounds,calc,positioning,fit,petri,plotmarks}
\usetikzlibrary{arrows}

\usepackage{enumerate}
\usepackage{paralist}

\newcommand*{\affaddr}[1]{#1} 
\newcommand*{\affmark}[1][*]{\textsuperscript{#1}}

\global\long\def\bX{\mathbf{X}}

\global\long\def\bY{\mathbf{Y}}

\global\long\def\bZ{\mathbf{Z}}

\global\long\def\bA{\mathbf{A}}

\global\long\def\bH{\mathbf{H}}

\global\long\def\bW{\mathbf{W}}

\global\long\def\bM{\mathbf{M}}

\global\long\def\bS{\mathbf{S}}

\global\long\def\bSigma{\boldsymbol{\Sigma}}
\global\long\def\bgamma{\boldsymbol{\gamma}}

\global\long\def\btheta{\boldsymbol{\theta}}
\global\long\def\bphi{\boldsymbol{\phi}}

\global\long\def\bPi{\boldsymbol{\Pi}}
\global\long\def\bpi{\boldsymbol{\pi}}
\global\long\def\bLambda{\boldsymbol{\Lambda}}

\global\long\def\bTheta{\boldsymbol{\Theta}}

\global\long\def\balpha{\boldsymbol{\alpha}}
\global\long\def\bmu{\boldsymbol{\mu}}
\global\long\def\bzeta{\boldsymbol{\zeta}}

\newcommand{\indep}{\rotatebox[origin=c]{90}{$\models$}}

\usepackage{xr}
\makeatletter
\newcommand*{\addFileDependency}[1]{
  \typeout{(#1)}
  \@addtofilelist{#1}
  \IfFileExists{#1}{}{\typeout{No file #1.}}
}
\makeatother
 
\newcommand*{\myexternaldocument}[1]{%
    \externaldocument{#1}%
    \addFileDependency{#1.tex}%
    \addFileDependency{#1.aux}%
}

\myexternaldocument{}

\title{Mediation Analysis with Graph Mediator}

\author{%
    Yixi Xu\affmark[1] and Yi Zhao\affmark[1] \\
    \affaddr{\affmark[1]Department of Biostatistics and Health Data Science \\
    Indiana University School of Medicine} \\
}

\date{}

\providecommand{\keywords}[1]
{
  {\small	
  \textbf{Keywords:} #1 }
}

\begin{document}

\maketitle

\thispagestyle{empty}

\begin{abstract}
This study introduces a mediation analysis framework when the mediator is a graph. A Gaussian covariance graph model is assumed for graph representation. Causal estimands and assumptions are discussed under this representation. With a covariance matrix as the mediator, parametric mediation models are imposed based on matrix decomposition. Assuming Gaussian random errors, likelihood-based estimators are introduced to simultaneously identify the decomposition and causal parameters. An efficient computational algorithm is proposed and asymptotic properties of the estimators are investigated. Via simulation studies, the performance of the proposed approach is evaluated. Applying to a resting-state fMRI study, a brain network is identified within which functional connectivity mediates the sex difference in the performance of a motor task.
\end{abstract}



\clearpage
\setcounter{page}{1}

\section{Introduction}
\label{sec:intro}

Mediation analysis has been widely used in clinical and biomedical studies to delineate the intermediate effect of a third variable, called mediator, in the causal pathway between the exposure/treatment and the outcome. It helps dissect the underlying causal mechanism by decomposing the effect of the exposure on the outcome into the part through the mediator and the part not through the mediator. Since the introduction of the classic Baron and Kenny framework~\citep{baron1986moderator}, mediation analysis has been extensively studied over decades, including under the causal inference framework~\cite[see review articles by][among others]{imai2010identification,vanderweele2015explanation,vanderweele2016mediation}. With the advances of biological technologies, mediation analysis has been extended to study cases of high-dimensional mediator~\cite[such as][among others]{huang2016hypothesis,chen2017high,derkach2019high,zhao2022pathway} and complex data output, including functional data~\citep{lindquist2012functional,zhao2018functional,zeng2021causal}, time series data~\citep{gu2014state,zhao2019granger}, network mediator~\citep{zhao2022bayesian}, image mediator~\citep{jiang2023causal,chen2023causal}, and so on.
In this study, we introduce a framework considering a graph as the mediator. 

Our study is motivated by resting-state functional magnetic resonance imaging (fMRI) experiments. A primary interest in resting-state fMRI is to portray brain coactivation patterns, the so-called brain functional connectivity or connectome, when the participant is at rest without any external stimulus. These coactivation patterns are captured by the covariance matrices (or correlation matrices after standardizing the data) of the fMRI signals extracted from brain voxels or regions of interest (ROI), where each ROI is a group of voxels defined by a chosen brain parcellation atlas~\citep{friston2011functional}. 
Considering a cognitive behavior, it is also of great interest to understand the brain mechanism related to the divergence in the behavior under various exposure/treatment conditions. Formulating it into a mediation analysis framework, brain functional connectivity is considered as the mediator between the exposure and behavior. Straightforward approaches include pairwise mediation analysis, where mediation analysis is repeated for each pair of functional connectivity followed by a multiple-testing correction, and high-dimensional mediation analysis, where the mediator is the vectorization of the upper triangular portion of the connectivity matrix. The former ignores the dependence between the correlations and suffers from power deficiency due to multiplicity and the latter disregards the structure property and the positive definiteness of a covariance matrix.

\begin{figure}
  \begin{center}
    \includegraphics[width=0.4\textwidth,page=2,trim={0.5cm 1.5cm 0.25cm 1.5cm},clip]{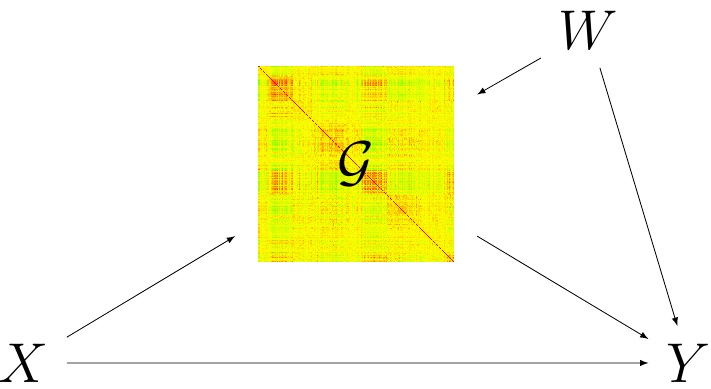}
  \end{center}
  \caption{\label{fig:dag}A conceptual mediation diagram with graph mediator ($\mathcal{G}$). $X$: exposure; $Y$: outcome; $W$: confounding factor.}
\end{figure}

In general, resting-state neuronal fluctuations measured by fMRI can be considered as Gaussian processes~\citep{lindquist2008statistical}. The functional connectivity matrix can thus be modeled via a \textit{Gaussian covariance graph model}. As a subclass of graphical models, the Gaussian covariance graph model defines a correspondence between the graph and the correlation pattern embedded in the covariance matrix~\citep{richardson2002ancestral,chaudhuri2007estimation}. A missing connection between two nodes in the graph is in concordance with zero correlation between the two corresponding variables in the covariance matrix, which also coincides with the marginal independence.
In this study, we will consider such a connectivity graph as the mediator to articulate the intermediate effect of brain connectome on the pathway between exposure and cognitive behavior. Figure~\ref{fig:dag} presents a conceptual diagram, where the graph is the mediator (denoted as $\mathcal{G}$). Closely related to this concept, \citet{zhao2022bayesian} recently proposed a Bayesian network mediation analysis. In their framework, stochastic block models (SBMs) were employed on individual elements in the adjacency matrix of brain regions and the vectorized model parameters were treated as the latent mediators to integratively impose feature selection and effect estimation. As the network mediator was represented by the adjacency matrix, only symmetry was imposed on the matrix structure and less stringent conditions were required for modeling.
In this study, instead of assuming underlying models on the adjacency matrix, it is proposed to directly model the covariance matrix as the mediator. It extends the linear structural equation modeling (LSEM) framework with a network mediator and at the same time preserves the positive definite property of a covariance matrix in parsimonious modeling.
Here, we would also like to distinguish the proposed framework from the case of image mediator considered by \citet{jiang2023causal} and \citet{chen2023causal}, where an image is (multidimensional) scalar output with spatial information.

Regression models for covariance matrix outcomes have been long before studied to capture the heterogeneity in the covariance structure across individuals/populations. In order to maintain the positive definiteness in modeling, example parametric approaches include modeling the covariance matrix as a linear combination of symmetric matrices~\citep{anderson1973asymptotically}, modeling the logarithmic transformation of matrix elements~\citep{chiu1996matrix}, modeling the covariance matrix via a quadratic link function~\citep{hoff2012covariance,seiler2017multivariate} or a linear combination of similarity matrices~\citep{zou2017covariance} of the covariates, and decomposition approaches based on the common diagonalization assumption via either eigendecomposition~\citep{flury1984common,boik2002spectral,hoff2009hierarchical,franks2019shared,zhao2021covariate} or Cholesky decomposition~\citep{pourahmadi2007simultaneous}.
In this study, we adapt the covariance regression framework introduced by \citet{zhao2021covariate} to mediation analysis. In their framework, orthogonal common linear projections were assumed on the covariance matrices and data heteroscedasticity in the projection space was revealed via a log-linear model. The advantages are to preserve the positive definiteness of the covariance matrices, offer high flexibility in modeling, and enable a network-level interpretation, where each projection can be considered a subnetwork after proper thresholding or sparsifying.
Integrating this framework with LSEM, common linear projections are assumed on the covariance matrices and the logarithmic transformation of data variance in the projection space is considered as the mediator measurement in the structural equation models. The objective is to simultaneously identify the linear projections and the causal parameters.
Contributions of this study include
\begin{inparaenum}[(i)]
  \item to enrich mediation analysis literature with graph mediator;
  \item to extend covariance regression modeling to SEM framework;
  \item and to offer a better understanding of brain mechanisms when studying the exposure effect on cognitive behaviors.
\end{inparaenum}

The rest of the manuscript is organized as follows. Section~\ref{sec:model} discusses the causal definitions and assumptions and the proposed parametric mediation model when the mediator is a graph. Likelihood-based estimators are proposed for estimating model parameters, an efficient algorithm is introduced for computation, and a resampling approach is considered for inference. Asymptotic properties of the estimators are investigated under regularity conditions. Via simulation studies in Section~\ref{sec:sim}, the performance of the proposed framework is evaluated and compared with competing approaches. Section~\ref{sec:application} applies the proposed mediation analysis to data acquired from the National Consortium on Alcohol and NeuroDevelopment in Adolescence (NCANDA) to dissect the mediation effect of brain functional connectome on sex difference in the performance of a motor task. Section~\ref{sec:discussion} summarizes the manuscript with a discussion.

\section{Model and Methods}
\label{sec:model}

\subsection{Causal definitions and assumptions}
\label{sub:causal_def}

Let $\mathcal{G}(x)=(\mathcal{V},\mathcal{E}(x))$ denote the graph object when the exposure is at level $x$, where $\mathcal{V}$ is the set of vertices/nodes and $\mathcal{E}(x)$ is the set of edges under the exposure level. 
Denote $\mathscr{G}=\{\mathcal{G}\}$ as the set of such graph objects.
A type of graph, \textit{covariance graph}, is considered in this study, which can be represented by a covariance matrix embedded with association structure between vertices~\citep{chaudhuri2007estimation}. 
Denote $\bM(x)=(M_{1}(x),\dots,M_{p}(x))^\top\in\mathbb{R}^{p}$ as the potential outcome of the $p$-dimensional random variables in the vertex set $\mathcal{V}=\{1,\dots,p\}$ and $\bSigma(x)=(\sigma_{jk}(x))\in\mathbb{R}^{p\times p}$ as the corresponding covariance matrix when the exposure is $x$. As the study focus is on the covariance matrix, without loss of generality, the variable means are assumed to be zero. Assuming $\bM(x)$ follows a multivariate Gaussian distribution, a \textit{Gaussian covariance graph model} is considered, where a missing edge between two vertices is equivalent to the marginal independence between the corresponding random variables~\citep{edwards2012introduction}: for $\forall~j,k\in\mathcal{V}$ and $j\neq k$,
\begin{equation}
  (j,k)\notin\mathcal{E}(x) \quad \Leftrightarrow \quad M_{j}(x)~\indep~M_{k}(x) \quad \Leftrightarrow \quad \sigma_{jk}(x)=\sigma_{kj}(x)=0.
\end{equation}
Let $Y(x,\mathcal{G}(x))\in\mathbb{R}$ denote the potential outcome of $Y$ when the exposure is $x$ and the mediator graph is $\mathcal{G}(x)$. Let $\mathscr{X}$ denote the possible exposure values. Assume a binary exposure, $\mathscr{X}=\{0,1\}$. The following defines the average total effect (ATE) of the exposure on the outcome.
\begin{equation}\label{eq:TE}
  \tau_{\text{ATE}} = \mathbb{E}\left\{Y(1,\mathcal{G}(1))-Y(0,\mathcal{G}(0))\right\}.
\end{equation}
With a graph mediator, the above total effect can be decomposed into direct and indirect effects. Extending the framework in \citet{imai2010identification}, the following defines the average indirect effect (AIE) under exposure level $x$ and the average direct effect (ADE) when the mediator graph is at level $\mathcal{G}(x)$.
\begin{eqnarray}
  \tau_{\text{AIE}}(x) &=& \mathbb{E}\left\{Y(x,\mathcal{G}(1))-Y(x,\mathcal{G}(0))\right\}, \quad x=0,1; \label{eq:IE} \\
  \tau_{\text{ADE}}(x) &=& \mathbb{E}\left\{Y(1,\mathcal{G}(x))-Y(0,\mathcal{G}(x))\right\}, \quad x=0,1. \label{eq:DE}
\end{eqnarray}
The AIE quantifies the difference between the potential outcome corresponding to the graph of $\mathcal{G}(1)$ and that under $\mathcal{G}(0)$ fixing the exposure at level $x$. It is also called the \textit{natural indirect effect}~\citep{pearl2001direct} or the \textit{pure indirect effect} for $\tau_{\text{AIE}}(0)$ and \textit{total indirect effect} for $\tau_{\text{AIE}}(1)$~\citep{robins1992identifiability}. The ADE quantifies the exposure effect on the outcome not through the mediator. It is easy to show that the ATE is the sum of AIE and ADE,
\begin{equation}
  \tau_{\text{ATE}}=\tau_{\text{AIE}}(x)+\tau_{\text{ADE}}(1-x), \quad x=0,1.
\end{equation}

The following discusses the identification assumptions of the causal estimands defined above. For mediation analysis, the identification assumptions have been extensively discussed under various settings in the literature~\cite[examples include][and so on]{robins1992identifiability,pearl2001direct,imai2010identification,vanderweele2015explanation}. In this study, we extend the assumptions in \citet{imai2010identification} to the scenario of a graph mediator.
\begin{description}
  \item[Assumption A1 (SUTVA)] Stable unit treatment value assumption.

  \item[Assumption A2 (Positivity)] Let $X$ denote the actual treatment assignment, then $\mathbb{P}(X=x)>0$ for $\forall~x\in\mathscr{X}$.

  \item[Assumption A3 (Ignorability)] Let $Y(x,\boldsymbol{\mathcal{G}})$ denote the potential outcome of $Y$ under the exposure level $x$ and the mediator graph taking the value of $\boldsymbol{\mathcal{G}}\in\mathscr{G}$ and $\bW\in\mathbb{R}^{q}$ denote a vector of $q$-dimensional observed confounding factors.
    \begin{equation}
      \left\{Y(1,\boldsymbol{\mathcal{G}}),Y(0,\boldsymbol{\mathcal{G}}),\bM(1),\bM(0)\right\}~\indep~X~|~\bW,
    \end{equation}
  for any $\boldsymbol{\mathcal{G}}\in\mathscr{G}$.

  \item[Assumption A4 (Sequential ignorability)]
    \begin{equation}
      Y(x,\boldsymbol{\mathcal{G}})~\indep~\bM(x)~|~X=x,\bW,
    \end{equation}
    for any $x\in\mathscr{X}$ and $\boldsymbol{\mathcal{G}}\in\mathscr{G}$.

  \item[Assumption A5 (Parametric LSEM)] The following linear structural equation models (LSEMs) with independent errors are assumed.
    \begin{eqnarray}
      \log(\btheta^\top\bSigma(x)\btheta) &=& \alpha_{0}+\alpha x+\bW^\top\bphi_{1}+\eta, \label{eq:model_M} \\
      Y(x,\boldsymbol{\mathcal{G}}) &=& \gamma_{0}+\gamma x+\beta\log(\btheta^\top\bSigma\btheta) +\bW^\top\bphi_{2}+\epsilon, \label{eq:model_Y}
    \end{eqnarray}
    where $\btheta\in\mathbb{R}^{p}$ is a projection vector with unit $L_{2}$-norm (i.e. $\|\btheta\|_{2}=1$), $\{\alpha_{0},\alpha,\bphi_{1},\gamma_{0},\gamma,\beta,\bphi_{2}\}$ are model coefficients, and $\eta$ and $\epsilon$ are independent model errors with mean zero.
\end{description}

The SUTVA assumption~\citep{rubin1980randomization} assumes that the treatment assignment regime remains the same across units and the outcome of one unit is not influenced by the treatment assignment of other units. In the application, this no interference assumption holds as the imaging scan and cognition evaluation were performed for each individual independently. The positivity assumption is standard in causal inference literature when possible treatment assignments are finite. Assumption A3 assumes that there is no unmeasured mediator-exposure or outcome-exposure confounding given the observed covariates. In the application study, sex is considered as the preceding exposure, which is assumed to be naturally randomized among the population of the same age. 
Assumption A4 assumes the conditional independence between the potential outcome of $Y$ and the potential outcome of $\bM$ under the so-called single-world intervention graphs~\citep{richardson2013single}, which is a relaxation of the cross-world independence assumption~\citep{robins2010alternative}. Assumption 5 assumes parametric linear structural equation models, where $\boldsymbol{\mathcal{G}}$ and $\bSigma$ have a one-to-one correspondence under the Gaussian covariance graph model. With Assumptions A3--A5, it is sufficient to identify the ADE and AIE~\citep{andrews2020insights}. 
Considering a Gaussian covariance graph model, the covariance matrix ($\bSigma$) has a one-to-one correspondence with the graph object ($\mathcal{G}$). Under Assumption A5, the potential outcome of $Y$ under a multiple-worlds model can be expressed as
\begin{eqnarray*}
  Y(x,\mathcal{G}(x')) &=& \gamma_{0}+\gamma x+\beta(\alpha_{0}+\alpha x'+\bW^\top\bphi_{1}+\eta)+\bW^\top\bphi_{2}+\epsilon \\
  &=& \gamma x+\alpha\beta x' +(\gamma_{0}+\alpha_{0}\beta+\beta\bW^\top\bphi_{1}+\bW^\top\bphi_{2})+\beta\eta+\epsilon. 
\end{eqnarray*}
The following theorem gives the parametric representation of the causal estimands.
\begin{theorem}\label{thm:causal_estimand}
  Under Assumptions A1--A5,
  \begin{eqnarray*}
    && \tau_{\mathrm{ATE}}=\mathbb{E}\left\{Y(1,\mathcal{G}(1))-Y(0,\mathcal{G}(0))\right\}=\gamma+\alpha\beta; \\
    && \tau_{\mathrm{AIE}}(x)=\mathbb{E}\left\{Y(x,\mathcal{G}(1))-Y(x,\mathcal{G}(0))\right\}=\alpha\beta, \quad \text{for } x=0,1; \\
    && \tau_{\mathrm{ADE}}(x)=\mathbb{E}\left\{Y(1,\mathcal{G}(x))-Y(0,\mathcal{G}(x))\right\}=\gamma, \quad \text{for } x=0,1.
  \end{eqnarray*}
\end{theorem}

\subsection{Method}
\label{sub:method}

This section introduces an approach to estimate model parameters in~\eqref{eq:model_M} and~\eqref{eq:model_Y} and draw inference based on resampling techniques. The parameter set includes not only the model coefficients, but also the projection vector, $\btheta$. In addition, the covariance matrices, $\bSigma$'s, are not directly observable. Rather, the observed data is the realization of the vertices. Let $\bM_{it}=(M_{it1},\dots,M_{itp})^\top\in\mathbb{R}^{p}$ denote the $t$th replicate of the $p$ vertices acquired from unit $i$, for $t=1,\dots,T_{i}$ and $i=1,\dots,n$, where $T_{i}$ is the number of realizations in unit $i$ and $n$ is the number of units. Let $X_{i}$ denote the actual treatment assignment, $Y_{i}$ denote the observed outcome, and $\bW_{i}\in\mathbb{R}^{q}$ denote the $q$-dimensional observed confounding factors of unit $i$. Assume the model errors, $\eta_{i}$ and $\epsilon_{i}$, are normally distributed with mean zero and variances $\pi^{2}$ and $\sigma^{2}$, respectively. Models~\eqref{eq:model_M} and~\eqref{eq:model_Y} can be viewed as a multilevel model. Thus, it is proposed to estimate the parameters by maximizing the hierarchical likelihood, which has been shown to be asymptotically equivalent to maximizing the marginal likelihood~\citep{lee1996hierarchical}. To simplify the notations, let
\[
  \bX_{i}=\begin{pmatrix}
    X_{i}, \\
    \bW_{i}
  \end{pmatrix}\in\mathbb{R}^{q+1}, ~\balpha=\begin{pmatrix}
    \alpha \\
    \bphi_{1}
  \end{pmatrix}\in\mathbb{R}^{q+1}, ~\text{and } \bgamma=\begin{pmatrix}
    \gamma \\
    \bphi_{2}
  \end{pmatrix}\in\mathbb{R}^{q+1}.
\]
The following gives the negative hierarchical likelihood (with a constant difference).
\begin{eqnarray}\label{eq:hlikelihood}
  \ell &=& \sum_{i=1}^{n}\frac{T_{i}}{2}\left\{(\alpha_{0i}+\bX_{i}^\top\balpha)+(\btheta^\top\bS_{i}\btheta)\exp(-\alpha_{0i}-\bX_{i}^\top\balpha)\right\} \\
  && +\sum_{i=1}^{n}\frac{1}{2}\left\{\log\sigma^{2}+\frac{1}{\sigma^{2}}(Y_{i}-\gamma_{0}-\bX_{i}^\top\bgamma-\beta\log(\btheta^\top\bSigma_{i}\btheta))^{2}\right\}+\sum_{i=1}^{n}\frac{1}{2}\left\{\log\pi^{2}+\frac{1}{\pi^{2}}(\alpha_{0i}-\alpha_{0})^{2}\right\}, \nonumber
\end{eqnarray}
where $\alpha_{0i}=\alpha_{0}+\eta_{i}$ and $\bS_{i}=T_{i}^{-1}\sum_{t=1}^{T_{i}}\bM_{it}\bM_{it}^\top$. The first component in the likelihood corresponds to the conditional likelihood of $\bM_{it}$ given $\alpha_{0i}$ and $\bX_{i}$ under model~\eqref{eq:model_M}; the second component corresponds to the conditional likelihood of $Y_{i}$ given $\alpha_{0i}$, $\bX_{i}$, and the graph, $\bSigma_{i}$, under model~\eqref{eq:model_Y}; and the third component corresponds to the likelihood of the random intercept, $\alpha_{0i}$. For the second component in~\eqref{eq:hlikelihood}, the covariance matrices, $\bSigma_{i}$'s, are not directly observed. A sample counterpart needs to be introduced and utilized instead in practice for estimation. For fixed $p$, denote $\hat{\bSigma}_{i}$ as a consistent estimator of $\bSigma_{i}$. For example, the sample covariance matrix, $\bS_{i}$, is a consistent estimator of $\bSigma_{i}$. Denote $\hat{\ell}$ as the sample counterpart of $\ell$ by replacing $\bSigma_{i}$ with $\hat{\bSigma}_{i}$.
\begin{eqnarray}\label{eq:hlikelihood_est}
  \hat{\ell} &=& \sum_{i=1}^{n}\frac{T_{i}}{2}\left\{(\alpha_{0i}+\bX_{i}^\top\balpha)+(\btheta^\top\bS_{i}\btheta)\exp(-\alpha_{0i}-\bX_{i}^\top\balpha)\right\} \\
  && +\sum_{i=1}^{n}\frac{1}{2}\left\{\log\sigma^{2}+\frac{1}{\sigma^{2}}(Y_{i}-\gamma_{0}-\bX_{i}^\top\bgamma-\beta\log(\btheta^\top\hat{\bSigma}_{i}\btheta))^{2}\right\}+\sum_{i=1}^{n}\frac{1}{2}\left\{\log\pi^{2}+\frac{1}{\pi^{2}}(\alpha_{0i}-\alpha_{0})^{2}\right\}. \nonumber
\end{eqnarray}
\begin{lemma}\label{lemma:hlikelihood_const}
  For fixed $p$ and $n$, assume $\hat{\bSigma}_{i}$ is a consistent estimator of $\bSigma_{i}$ as $\min_{i}T_{i}\rightarrow\infty$, for $\forall~i\in\{1,\dots,n\}$. Then,
  \[
    \hat{\ell}\rightarrow\ell, \quad \text{as } \min_{i}T_{i}\rightarrow\infty.
  \]
\end{lemma}
\noindent With Lemma~\ref{lemma:hlikelihood_const}, it is proposed to estimate model parameters by minimizing $\hat{\ell}$. 
A coordinate descent algorithm is considered to solve for solution over the parameter set $\bTheta=(\btheta,\alpha_{0i},\alpha_{0},\balpha,\gamma_{0},\bgamma,\beta,\pi^{2},\sigma^{2})$. For $\btheta$, a constraint is required to identify a unique solution. The following optimization problem is considered.
\begin{eqnarray}
  \text{minimize} && \hat{\ell}, \nonumber \\
  \text{such that} && \btheta^\top\bH\btheta=1, \label{eq:obj_func}
\end{eqnarray}
where $\bH\in\mathbb{R}^{p\times p}$ is a positive definite matrix. The choice of $\bH$ can be subject dependent. Similar to the choice in PCA or CCA, one can set $\bH$ to the $p$-dimensional identity matrix. As $\bM_{it}$'s are assumed to be normally distributed, one can set $\bH=\bar{\bS}=\sum_{i=1}^{n}\sum_{j=1}^{T_{i}}\bM_{it}\bM_{it}^\top/\sum_{i=1}^{n}T_{i}$ to incorporate distributional information of the data. This type of choice was also considered in the study of common PCA~\citep{krzanowski1984principal} and a covariance regression model by \citet{zhao2021covariate}.

Algorithm~\ref{alg:capmed} summarizes the optimization steps of problem~\eqref{eq:obj_func}. For parameters without an analytic solution, including $\alpha_{0i}$ and $\balpha$, the Newton-Raphson algorithm is employed to find the update. For $\{\alpha_{0},\gamma_{0},\bgamma,\beta,\pi^{2},\sigma^{2}\}$, explicit solutions can be obtained and utilized for update. For $\btheta$, with the constraint, the following gives the Lagrangian form.
\begin{equation}
  \mathcal{L}(\btheta,\lambda)=\frac{1}{2}\sum_{i=1}^{n}\left\{\btheta^\top(T_{i}U_{i}\bS_{i})\btheta+\frac{1}{\sigma^{2}}(V_{i}-\beta\log(\btheta^\top\hat{\bSigma}_{i}\btheta))^{2}\right\}-\lambda(\btheta^\top\bH\btheta-1),
\end{equation}
where $U_{i}=\exp(-\alpha_{0i}-\bX_{i}^\top\balpha)$, $V_{i}=Y_{i}-\gamma_{0}-\bX_{i}^\top\bgamma$, and $\lambda$ is the Lagrangian parameter. Take the partial derivative over $\btheta$ and set it to zero, it gives
\begin{equation}\label{eq:pL_theta}
  \frac{\partial\mathcal{L}(\btheta,\lambda)}{\partial\btheta}=\sum_{i=1}^{n}\left[(T_{i}U_{i}\bS_{i})\btheta-\frac{1}{\sigma^{2}}\left\{V_{i}-\beta\log(\btheta^\top\hat{\bSigma}_{i}\btheta)\right\}\frac{2\beta\hat{\bSigma}_{i}\btheta}{\btheta^\top\hat{\bSigma}_{i}\btheta}\right]-2\lambda\bH\btheta=\boldsymbol{\mathrm{0}}.
\end{equation}
The above equation is hard to derive the explicit form to solve for $\btheta$. It is observed that the quantity of $\btheta^\top\hat{\bSigma}_{i}\btheta$ is in both a logarithmic function and the denominator. To simplify the computation, it is proposed to plug in the value of $\btheta$ from the previous step $s$ into $\btheta^\top\hat{\bSigma}_{i}\btheta$ denoted as $\xi_{i}^{(s)}=\btheta^{(s)\top}\hat{\bSigma}_{i}\btheta^{(s)}$. Let
\[
  \bA^{(s)}=\sum_{i=1}^{n}\left\{T_{i}U_{i}^{(s)}\bS_{i}-\frac{2\beta^{(s)}(V_{i}^{(s)}-\beta^{(s)}\log\xi_{i}^{(s)})}{\sigma^{2(s)}\xi_{i}^{(s)}}\hat{\bSigma}_{i}\right\},
\]
where $U_{i}^{(s)}$, $V_{i}^{(s)}$, $\beta^{(s)}$, and $\sigma^{2(s)}$ are the solutions from the $s$th step. It is proposed to solve the following instead of the equation in~\eqref{eq:pL_theta},
\begin{equation}\label{eq:solve_theta}
  \bA^{(s)}\btheta-\lambda\bH\btheta=\boldsymbol{\mathrm{0}},
\end{equation}
where the solution $(\btheta,\lambda)$ is the pair of eigenvector and eigenvalue of $\bA^{(s)}$ with respect to $\bH$ that minimizes $\mathcal{L}(\btheta,\lambda)$. More details of the optimization algorithm is presented in Section~\ref{appendix:sec:alg_covreg} of the supplementary materials. After obtaining an estimate of model parameters, causal estimands are estimated following Theorem~\ref{thm:causal_estimand}.

\begin{algorithm}
  \caption{\label{alg:capmed}The optimization algorithm for problem~\eqref{eq:obj_func}.}
  \begin{algorithmic}[1]
    \INPUT $\{X_{i},\{\bM_{i1},\dots,\bM_{iT_{i}}),Y_{i},\bW_{i}~|~i=1,\dots,n\}$

    \State \textbf{initialization}: $(\btheta^{(0)},\alpha_{0i}^{(0)},\alpha_{0}^{(0)},\balpha^{(0)},\gamma_{0}^{(0)},\bgamma^{(0)},\beta^{(0)},\pi^{2(0)},\sigma^{2(0)})$

    \Repeat \; for iteration $s=0,1,2,\dots$

    \State \; update $\balpha$ and $\{\alpha_{0i}\}$ using the Newton-Raphson algorithm, denoted as $\balpha^{(s+1)}$ and $\{\alpha_{0i}^{(s+1)}\}$,

    \State \; update $(\alpha_{0},\gamma_{0},\bgamma,\beta,\pi^{2},\sigma^{2})$ with
      \[
        \alpha_{0}^{(s+1)}=\frac{1}{n}\sum_{i=1}^{n}\alpha_{0i}^{(s+1)}, \quad \pi^{2(s+1)}=\frac{1}{n}\sum_{i=1}^{n}\left(\alpha_{0i}^{(s+1)}-\alpha_{0}^{(s+1)}\right)^{2},
      \]
      \[
        \bmu^{(s+1)}=\left(\sum_{i=1}^{n}\bZ_{i}^{(s)}\bZ_{i}^{(s)\top}\right)^{-1}\left(\sum_{i=1}^{n}Y_{i}\bZ_{i}^{(s)}\right), \quad \sigma^{2(s+1)}=\frac{1}{n}\sum_{i=1}^{n}\left(Y_{i}-\bZ_{i}^{(s)\top}\bmu^{(s+1)}\right)^{2},
      \]
      where
      \[
        \bZ_{i}^{(s)}=\begin{pmatrix}
          1 \\
          \bX_{i} \\
          \log(\btheta^{(s)\top}\bS_{i}\btheta^{(s)})
        \end{pmatrix} \text{ and } \bmu^{(s+1)}=\begin{pmatrix}
          \gamma_{0}^{(s+1)} \\
          \bgamma^{(s+1)} \\
          \beta^{(s+1)}
        \end{pmatrix},
      \]

    \State \; update $\btheta$ by solving~\eqref{eq:solve_theta}, denoted as $\btheta^{(s+1)}$,

    \Until{the objective function in~\eqref{eq:obj_func} converges;}

    \State consider a random series of initializations, repeat Steps 1--6, and choose the estimates with the minimum objective value.

    \OUTPUT $(\hat{\btheta},\hat{\alpha}_{0i},\hat{\alpha}_{0},\hat{\balpha},\hat{\gamma}_{0},\hat{\bgamma},\hat{\beta},\hat{\pi}^{2},\hat{\sigma}^{2})$
  \end{algorithmic}
\end{algorithm}

Algorithm~\ref{alg:capmed} offers an approach to identify one mediation component based on the likelihood criterion. To identify higher-order components, it is proposed to remove the identified components from the data first and then replace the input in Algorithm~\ref{alg:capmed} with the new data to identify the next component. Let $\hat{\bTheta}^{(k)}=(\hat{\btheta}_{1},\dots,\hat{\btheta}_{k})\in\mathbb{R}^{p\times k}$ denote the first $k$ identified components. For $i=1,\dots,n$, set 
\begin{equation}
  \hat{\bM}_{i}^{(k+1)}=\bM_{i}-\bM_{i}\hat{\bTheta}^{(k)}\hat{\bTheta}^{(k)\top} \text{ and } \hat{Y}_{i}^{(k+1)}=Y_{i}-\sum_{j=1}^{k}\hat{\beta}_{j}\log(\hat{\btheta}_{j}^\top\hat{\bSigma}_{i}\hat{\btheta}_{j})
\end{equation}
as the new data, where $\bM_{i}=(\bM_{i1},\dots,\bM_{iT_{i}})^\top\in\mathbb{R}^{T_{i}\times p}$ is the mediator outcome of unit $i$ and $\hat{\beta}_{j}$ is the estimate of $\beta$ in the $j$th component, for $j=1,\dots,k$. $\hat{\bM}_{i}^{(k+1)}$ is created following a similar strategy as in the principal component analysis to identify orthogonal components, which is also employed in a recently introduced covariance regression model~\citep{zhao2021covariate}. It is also proposed to remove the identified mediation effect from the outcome in order to identify a new mediation component. In this sense, the mediation mechanism of the identified components are considered parallel. Fitting the mediation components marginally is then equivalent to fitting them jointly~\citep{vanderweele2015explanation}.
To determine the number of mediation components, a criterion called the average deviation from diagonality introduced in \citet{zhao2021covariate} is utilized, which is defined as
\begin{equation}\label{eq:DfD}
  \mathrm{DfD}(\hat{\bTheta}^{(k)})=\prod_{i=1}^{n}\left[\frac{\det\left\{\mathrm{diag}(\hat{\bTheta}^{(k)\top}\hat{\bSigma}_{i}\hat{\bTheta}^{(k)})\right\}}{\det(\hat{\bTheta}^{(k)\top}\hat{\bSigma}_{i}\hat{\bTheta}^{(k)})}\right]^{T_{i}/\sum_{i}T_{i}},
\end{equation}
where $\mathrm{diag}(\bA)$ is a diagonal matrix taking the diagonal elements in a square matrix $\bA$ and $\det(\bA)$ is the determinant of $\bA$. The above $\mathrm{DfD}$ metric achieves its minimum value of one when $\hat{\bTheta}^{(k)}$ commonly diagonalizes $\hat{\bSigma}_{i}$ for all $i\in\{1,\dots,n\}$. As $k$ increases, it gets more difficult to diagonalize all estimated covariance matrices and the value of the $\mathrm{DfD}$ metric may increase dramatically. The number of components can be then chosen by setting a threshold. For example, choose $k$ such that $\mathrm{DfD}(\hat{\bTheta}^{(k)})\leq 2$ as recommended in \citet{zhao2021covariate}.

\subsection{Inference}
\label{sub:inference}

To draw inference on the causal estimands, a resampling procedure is considered. Particularly, inference on the average indirect effect, which is denoted as the product of $\alpha$ and $\beta$ according to Theorem~\ref{thm:causal_estimand}, raises concerns. In general, even under the normality assumption, the distribution of the product of two estimates can be far from Gaussian in finite sample. Thus, the following nonparametric bootstrap procedure is introduced.
\begin{description}
  \item[Step 0.] Using all sample units, $k$ mediation components are identified and denote the estimated linear projections as $\hat{\btheta}_{j}$, for $j=1,\dots,k$.
  \item[Step 1.] Generate a bootstrap sample of size $n$ by sampling with replacement, denoted as \\ $\{X_{i}^{*},(\bM_{i1},\dots,\bM_{iT_{i}})^{*},Y_{i}^{*},\bW_{i}^{*}~|~i=1,\dots,n\}$.
  \item[Step 2.] For $j=1,\dots,k$, using the resampled data, estimate model coefficients and variances following Algorithm~\ref{alg:capmed} with $\btheta=\hat{\btheta}_{j}$.
  \item[Step 3.] Repeat Steps 1--2 for $B$ times.
  \item[Step 4.] Construct bootstrap confidence intervals for the causal estimands under a prespecified significance level. 
\end{description}
As the study focus is on the causal parameters, the resampling is conducted at the unit level. Once a unit is sampled, all mediator observations remain for estimation.
The above procedure is implemented fixing the $k$ identified projections. Thus, inference on $\btheta$ is unachievable. If one seeks to perform inference on $\btheta$ via a bootstrap procedure, it requires a matching step on the estimates across bootstrap samples as the number of components and the order of identifying the components may differ. In addition, the inference performance can be sensitive to the quality of the matching step and the metrics used for matching. On the other hand, considering the computation cost, it will be significantly inflated when adding the estimation of $\btheta$ and matching. Thus, we leave the inference on $\btheta$ to future research.

\subsection{Asymptotic properties}
\label{sub:asmp}

In this section, we study the asymptotic properties of the proposed estimator. We first discuss the regularity conditions on the graph mediator. As discussed in Section~\ref{sub:causal_def}, under the Gaussian covariance graph model, the graph and the covariance matrix have one-to-one correspondence. Considering the covariance matrix, $\bSigma_{i}$ (for $i=1,\dots,n$), it is assumed that it has the eigendecomposition of $\bSigma_{i}=\bPi_{i}\bLambda_{i}\bPi_{i}^\top$, where $\bPi_{i}=(\bpi_{i1},\dots,\bpi_{ip})\in\mathbb{R}^{p\times p}$ is an orthonormal matrix and $\bLambda_{i}=\mathrm{diag}\{\lambda_{i1},\dots,\lambda_{ip}\}\in\mathbb{R}^{p\times p}$ is a diagonal matrix of corresponding eigenvalues. Let $\bzeta_{it}=\bPi_{i}^\top\bM_{it}\in\mathbb{R}^{p}$, and then $\mathrm{Cov}(\bzeta_{it})=\bLambda_{i}$. The elements in $\bzeta_{it}$ are uncorrelated. Under the normality assumption, they are mutually independent. In addition, the following assumptions are imposed.
\begin{description}
  \item[Assumption B1] Let $T=\min_{i}T_{i}$. $p\ll T$ is fixed.
  \item[Assumption B2] The eigenvectors of $\bSigma_{i}$ are identical, that is $\bPi_{i}=\bPi$, for $i=1,\dots,n$.
  \item[Assumption B3] For $\forall~i=1,\dots,n$, there exists (at least) one column in $\bPi_{i}$ indexed by $k_{i}$, such that $\btheta=\bpi_{ik_{i}}$ and the parametric model assumption (Assumption A5) holds.
\end{description}

Assumption B1 assumes a low-dimensional scenario for the Gaussian covariance graph model. Under this assumption, the sample covariance matrices are well-conditioned and consistent estimators. Replace $\hat{\bSigma}_{i}$ with the sample covariance matrix, Lemma~\ref{lemma:hlikelihood_const} holds. Assumption B2 is a common-diagonalization assumption assuming all the covariance matrices have the same set of eigenvectors. However, the order of the eigenvectors may vary if following the descending order of the corresponding eigenvalues. Assumption B3 assumes the parametric models in Assumption A5 are correctly specified. Under these assumptions, one can choose the eigenvectors of $\bar{\bS}$ as the initial value of $\btheta$ in Algorithm~\ref{alg:capmed}, where $\bar{\bS}=\sum_{i=1}^{n}\sum_{j=1}^{T_{i}}\bM_{it}\bM_{it}^\top/\sum_{i=1}^{n}T_{i}$ is the average sample covariance across all units. The following proposition demonstrates the consistency of the proposed estimator.
\begin{proposition}\label{prop:est_asmp}
  Assume Assumptions B1--B3 hold. As $n,T\rightarrow\infty$, the proposed estimator of model parameters is asymptotically consistent.
\end{proposition}

\section{Simulation Study}
\label{sec:sim}

This section compares the performance of the proposed mediation framework with a competing approach via simulation studies. As no existing approach can be directly implemented for mediation analysis with a graph mediator, or when the mediator is a covariance matrix, an approach integrating the covariate assisted principal (CAP) regression~\citep{zhao2021covariate} and regression-based mediation analysis, named as \textbf{CAP-Med}, is considered as the competing method. The CAP-Med approach has two steps.
\begin{inparaenum}[(i)]
  \item Perform CAP analysis under Model~\eqref{eq:model_M} to identify the mediator components.
  \item For each identified component in Step (i), perform the regression-based mediation analysis as in \citet{imai2010identification}.
\end{inparaenum}
The proposed graph mediation approach is named as \textbf{GMed}.

Two simulation studies are conducted:
\begin{inparaenum}[(1)]
  \item to examine and compare the performance of GMed to CAP-Med; and
  \item to examine the robustness of GMed to model misspecification.
\end{inparaenum}
In both studies, common eigenstructure is assumed across covariance matrices and the covariance matrices are generated from the eigendecomposition, $\bSigma_{i}=\bPi\bLambda_{i}\bPi^\top$, where $\bPi=(\bpi_{1},\dots,\bpi_{p})\in\mathbb{R}^{p\times p}$ is an orthonormal matrix and $\bLambda_{i}=\mathrm{diag}\{\lambda_{i1},\dots,\lambda_{ip}\}\in\mathbb{R}^{p\times p}$ is a diagonal matrix of individual eigenvalues, for $i=1,\dots,n$. In Simulation (1), two components, the second component (D2) and the forth component (D4), are chosen to satisfy the model assumption, Assumption A5. In both components, the magnitude of model parameters are set to one and both the direct effect ($\tau_{\text{ADE}}$) and indirect effect ($\tau_{\text{AIE}}$) are with the value of one. The exposure ($X$) is generated from a Bernoulli distribution with probability $0.5$ of being one. The random errors, $\eta$ and $\epsilon$, are generated from normal distributions with mean zero and standard deviation $0.1$. For the rest dimensions, the eigenvalues are generated from a normal distribution with mean value exponentially decaying from $3$ to $-1$ and standard deviation $0.1$. With the generated covariance matrices, $\bM_{it}$ are generated from the multivariate normal distribution with mean zero and $Y_{i}$ is generated from Model~\eqref{eq:model_Y}. In this simulation study, no covariate ($W$) is considered. 
Two scenarios of data dimension are considered, $p=10$ and $p=50$, and the sample sizes are set to be $(n,T)=(500,100)$ and $(n,T)=(500,500)$, where the number of observations within each unit is set to be identical ($T_{i}=T$). The number of units ($n$) is set to $500$ to be comparable with the sample size in the application study in Section~\ref{sec:application}.
In Simulation (2), two covariates ($q=2$), $\bW=(W_{1},W_{2})^\top$, one continuous and one binary, are considered and the corresponding model coefficients with the magnitude of $0.5$. The continuous covariate ($W_{1}$) is generated from a normal distribution with mean zero and standard deviation $0.5$ and the binary covariate ($W_{2}$) is generated from a Bernoulli distribution with probability $0.5$ of being one. The rest model parameters are the same as in Simulation (1) and data are generated following~\eqref{eq:model_M} and~\eqref{eq:model_Y}. In this simulation, to evaluate the performance to model misspecification, the proposed GMed approach is applied ignoring the covariates, named as \textbf{GMed-Mis}. The result under the correctly specified model is denoted as \textbf{GMed}.
To evaluate the performance of identifying mediation components, the absolute value of the inner product of the estimated projection and the truth, denoted as $|\langle \hat{\btheta}, \bpi_{j}\rangle|$ ($j=2,4$), is used as a metric of similarity between two unit-norm vectors.
For both CAP and GMed approaches, the number of components is determined using the $\mathrm{DfD}\leq 2$ criterion.
Simulations are repeated for $200$ times.

\begin{figure}
  \begin{center}
    \subfloat[D2: $|\langle\hat{\btheta},\bpi_{j}\rangle|$]{\includegraphics[width=0.33\textwidth]{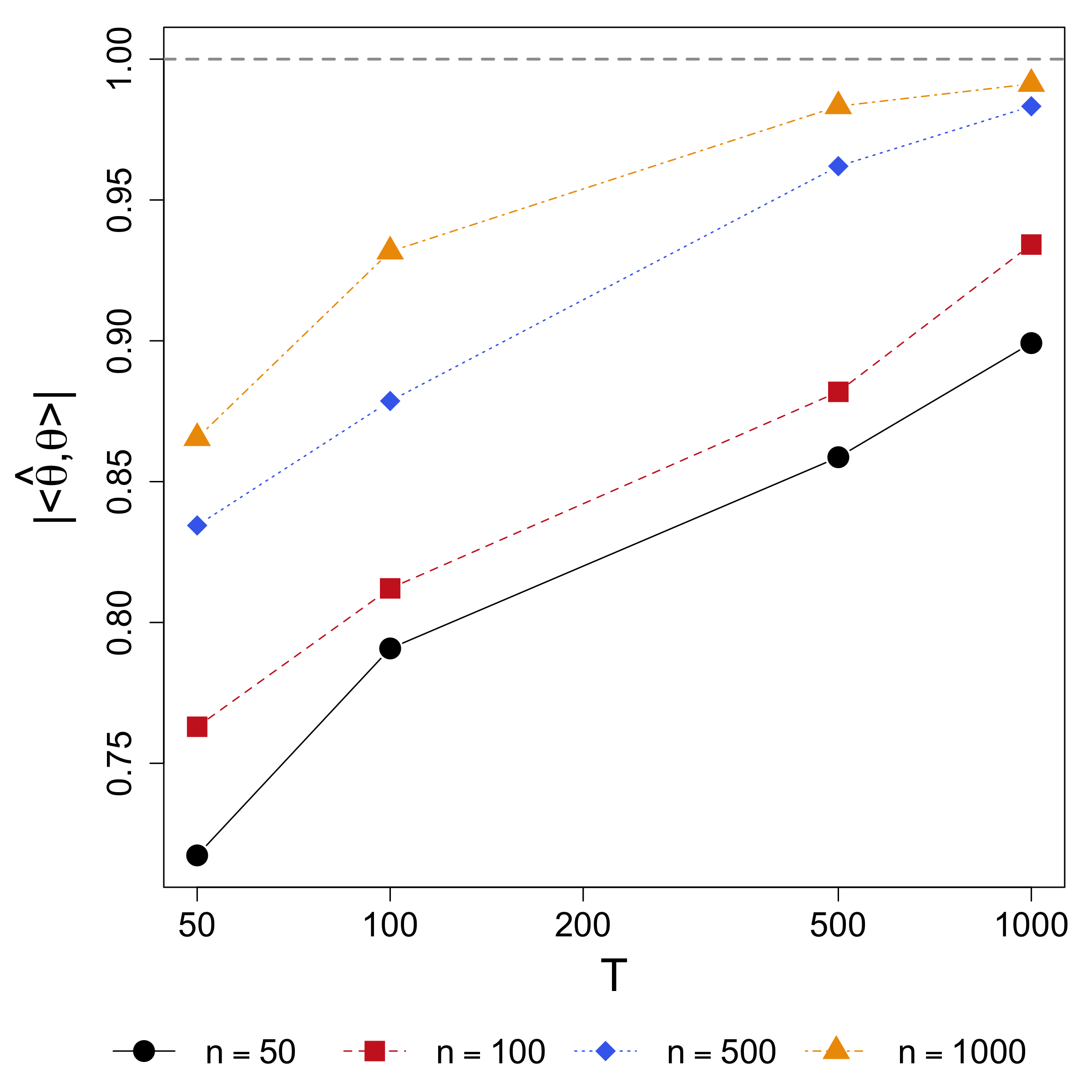}}
    \subfloat[D2: Bias of $\hat{\tau}_{\mathrm{AIE}}$]{\includegraphics[width=0.33\textwidth]{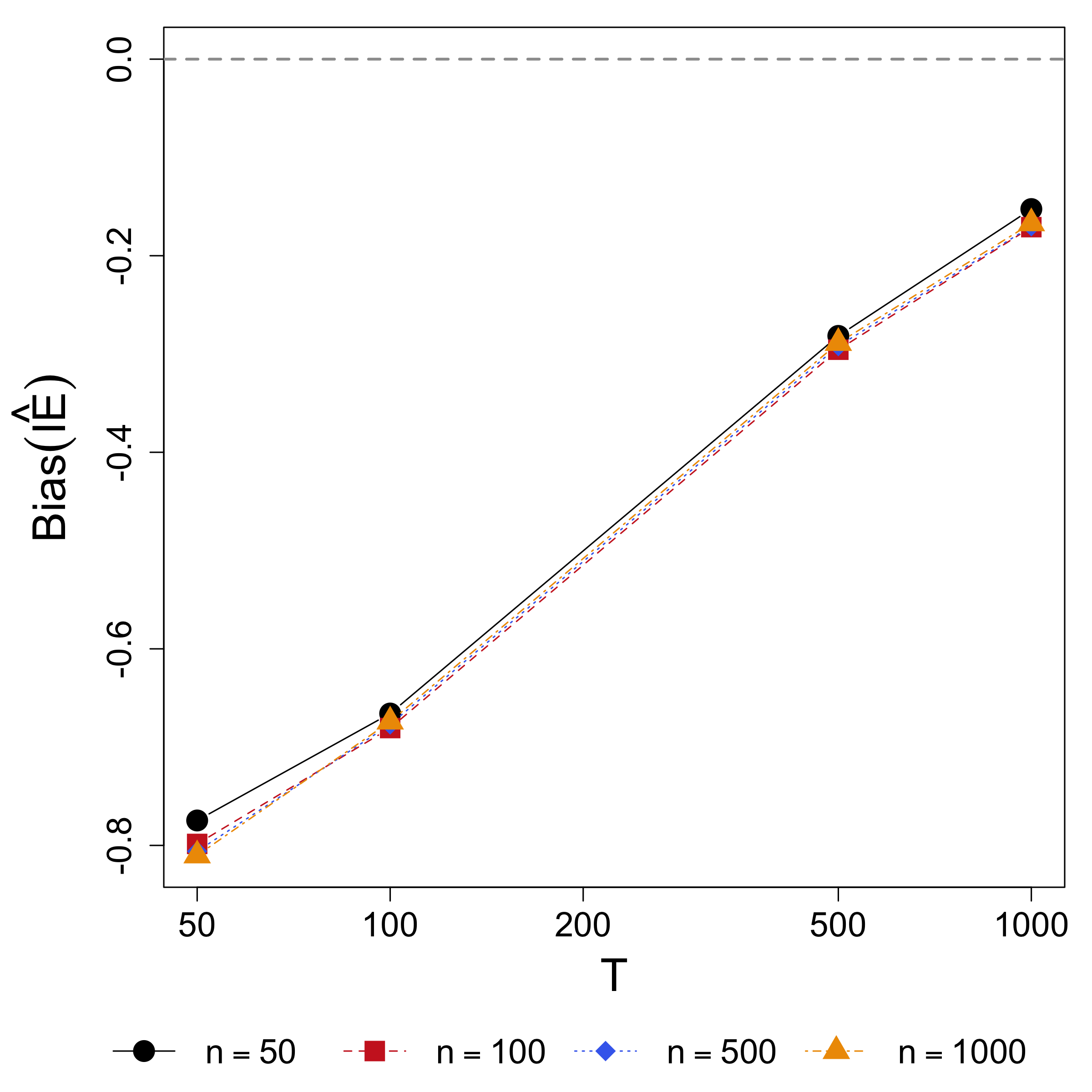}}
    \subfloat[D2: MSE of $\hat{\tau}_{\mathrm{AIE}}$]{\includegraphics[width=0.33\textwidth]{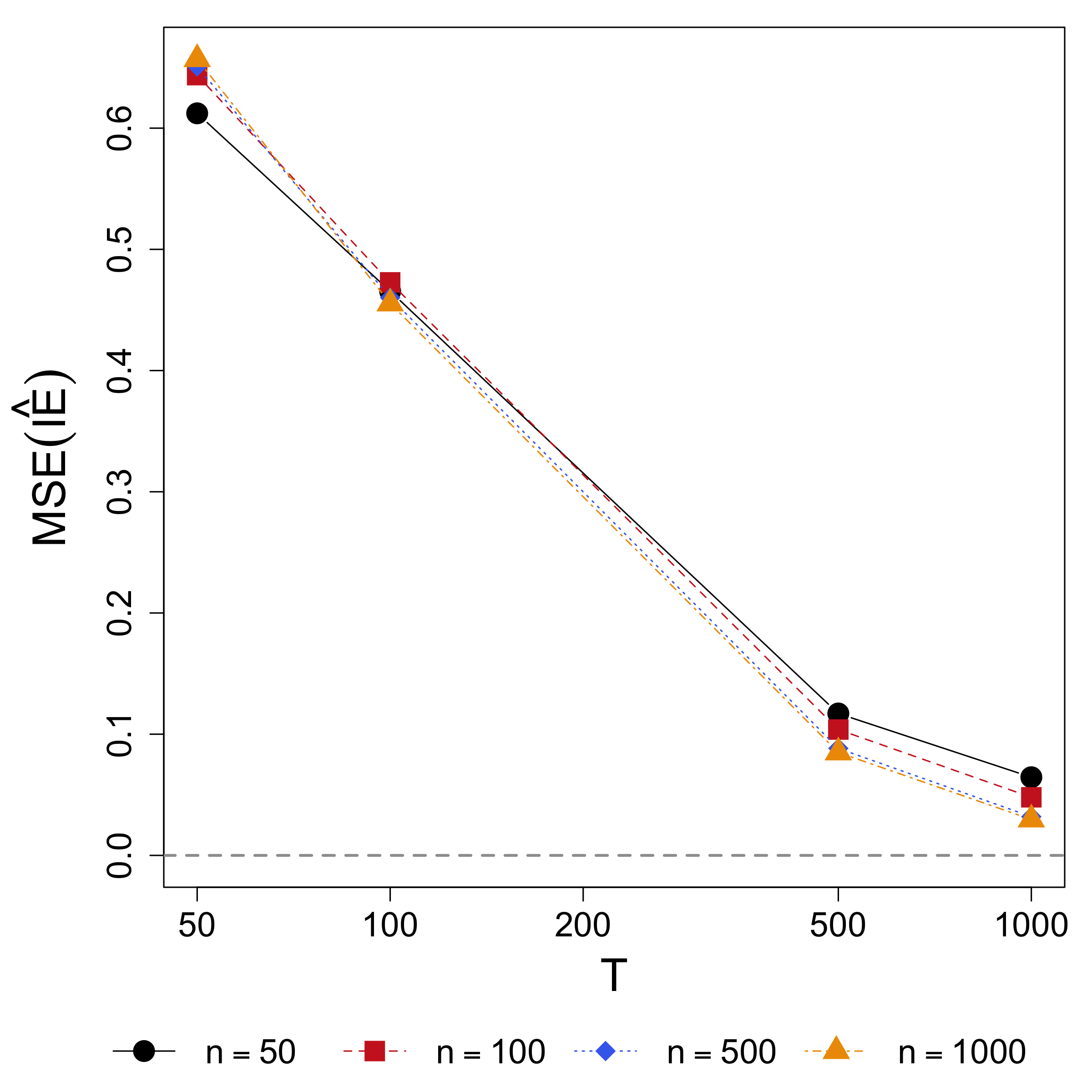}}

    \subfloat[D4: $|\langle\hat{\btheta},\bpi_{j}\rangle|$]{\includegraphics[width=0.33\textwidth]{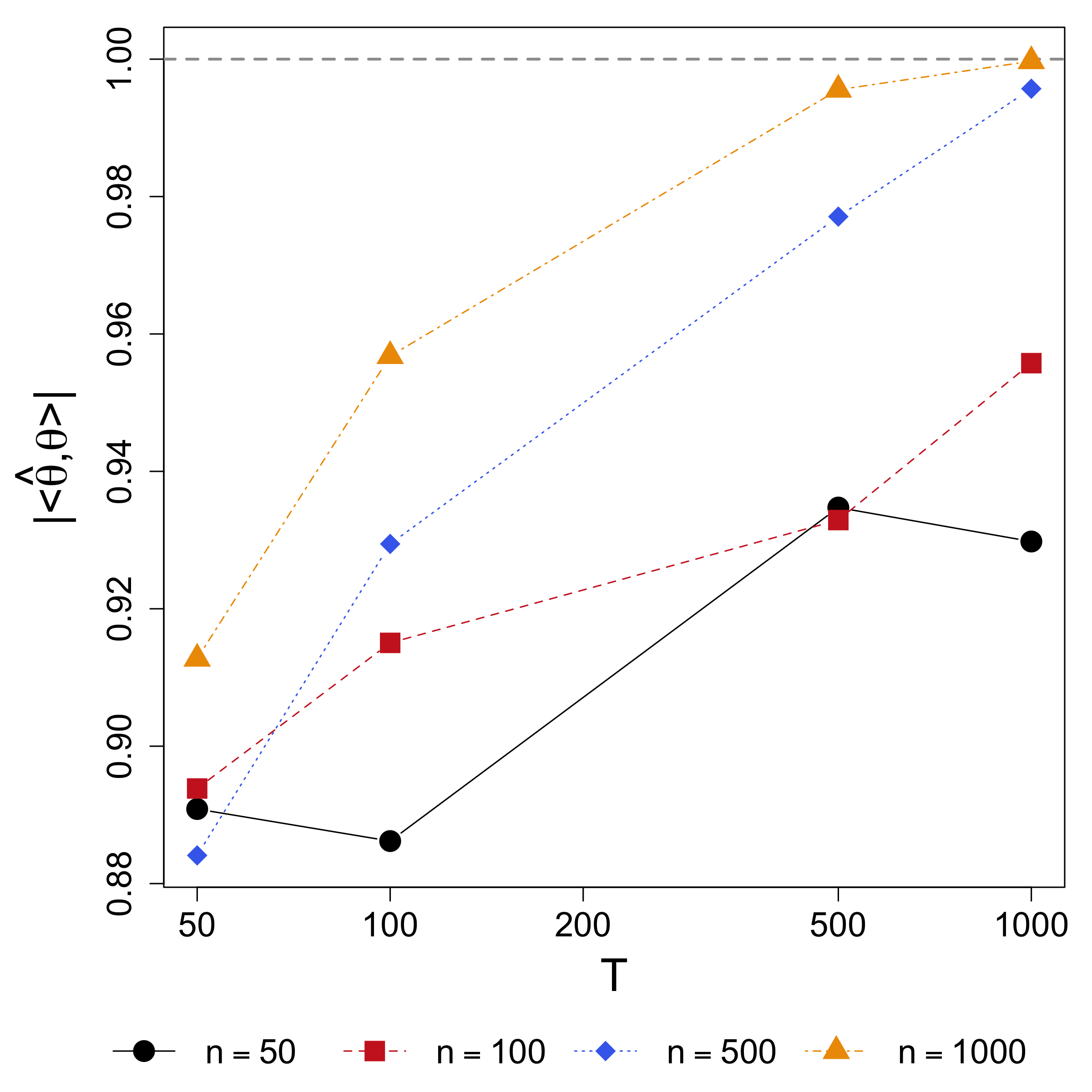}}
    \subfloat[D4: Bias of $\hat{\tau}_{\mathrm{AIE}}$]{\includegraphics[width=0.33\textwidth]{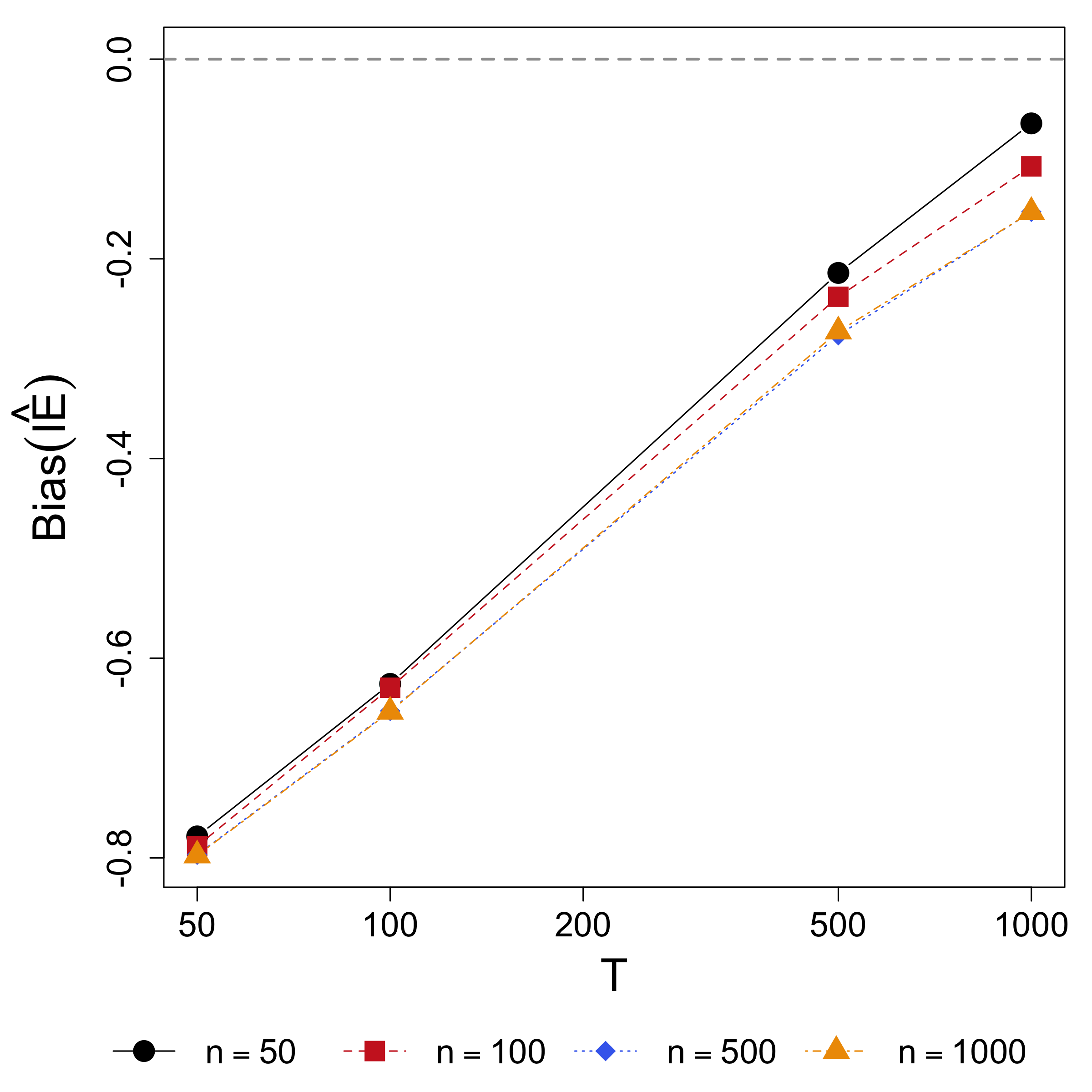}}
    \subfloat[D4: MSE of $\hat{\tau}_{\mathrm{AIE}}$]{\includegraphics[width=0.33\textwidth]{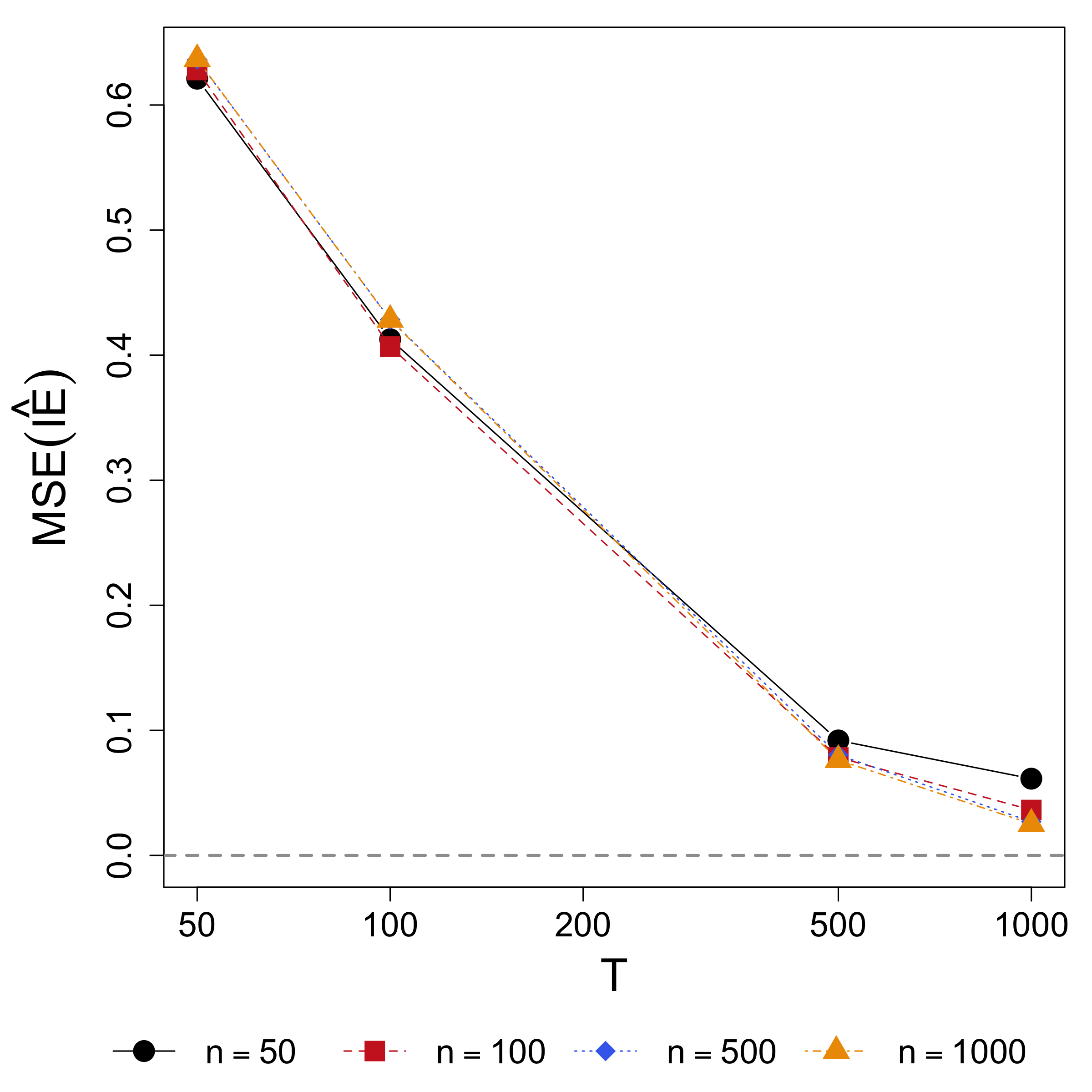}}
  \end{center}
  \caption{\label{fig:sim}Performance of estimating target mediation components ($\btheta$) and corresponding average indirect effect ($\tau_{\text{AIE}}$) over $200$ replicates in the simulation study as the sample size $(n,T)$ varies with $p=10$. MSE: mean squared error.}
\end{figure}

Table~\ref{tabel:sim} presents the performance of estimating mediation components and the corresponding average indirect effect. In Simulation (1), considering Model~\eqref{eq:model_M} only, the CAP approach shall achieve the optimal performance in estimating $\btheta$ and model coefficients as the method is likelihood-based and assumptions required are satisfied. The proposed approach yields comparable performance to the CAP-Med approach. As the number of observations within each unit increases, the performance of GMed in estimating $\btheta$ improves more significantly compared to the CAP-Med approach. In addition, the GMed approach performs consistently better in estimating $\btheta$ in the second mediation component (D4). This suggests that incorporating the outcome information in the estimating procedure helps improve performance.
To evaluate the finite sample performance of the GMed approach, Figure~\ref{fig:sim} presents the performance with $p=10$ and various sample size combinations. From the figure, as both the number of units and the number of observations within each unit increase, the similarity of $\hat{\btheta}$ to the truth converges to one and the bias and mean squared error (MSE) of the estimated average indirect effect converge to zero.

Simulation (2) aims to evaluate the robustness of GMed to model misspecification. Under the current setting, it also examines the robustness of GMed to the existence of an additive unmeasured mediator-outcome confounding not induced by the exposure. From Table~\ref{tabel:sim}, when the models are misspecified, the projections are still correctly identified though the estimate of the average indirect effect is off. This demonstrates the robustness of the GMed approach in estimating relevant mediation components. Utilizing this property, Section~\ref{appendix:sub:real_sens} of the supplementary materials suggests an approach for sensitivity analysis. 

\begin{table}
  \caption{\label{tabel:sim}Performance of estimating target mediation components ($\btheta$) and corresponding average indirect effect ($\tau_{\text{AIE}}$) over $200$ replicates in the simulation study. SE: standard error; MSE: mean squared error.}
  \begin{center}
    \begin{tabular}{c c c l c r c r r}
      \hline
      & & & & & \multicolumn{1}{c}{$\hat{\btheta}$} && \multicolumn{2}{c}{$\hat{\tau}_{\mathrm{AIE}}$} \\
      \cline{6-6}\cline{8-9}
      \multicolumn{1}{c}{\multirow{-2}{*}{Case}} & \multicolumn{1}{c}{\multirow{-2}{*}{$p$}} & \multicolumn{1}{c}{\multirow{-2}{*}{$(n,T)$}} & \multicolumn{1}{c}{\multirow{-2}{*}{Method}} & & \multicolumn{1}{c}{$|\langle\hat{\btheta},\bpi_{j}\rangle|$ (SE)} && \multicolumn{1}{c}{Bias} & \multicolumn{1}{c}{MSE} \\
      \hline
      & & & & D2 & $0.879$ ($0.113$) && $-0.677$ & $0.460$ \\
      & & & \multirow{-2}{*}{GMed} & D4 & $0.929$ ($0.140$) && $-0.653$ & $0.428$ \\
      \cline{4-9}
      & & & & D2 & $0.886$ ($0.138$) && $-0.664$ & $0.443$  \\
      & & \multirow{-4}{*}{$(500,100)$} & \multirow{-2}{*}{CAP-Med} & D4 & $0.905$ ($0.120$) && $-0.635$ & $0.406$ \\
      \cline{3-9}
      & & & & D2 & $0.962$ ($0.047$) && $-0.292$ & $0.088$ \\
      & & & \multirow{-2}{*}{GMed} & D4 & $0.977$ ($0.096$) && $-0.276$ & $0.080$ \\
      \cline{4-9}
      & & & & D2 & $0.911$ ($0.134$) && $-0.243$ & $0.067$ \\
      & \multirow{-8}{*}{$10$} & \multirow{-4}{*}{$(500,500)$} & \multirow{-2}{*}{CAP-Med} & D4 & $0.919$ ($0.121$) && $-0.129$ & $0.038$ \\
      \cline{2-9}
      & & & & D2 & $0.841$ ($0.101$) && $-0.665$ & $0.444$ \\
      & & & \multirow{-2}{*}{GMed} & D4 & $0.910$ ($0.142$) && $-0.659$ & $0.436$ \\
      \cline{4-9}
      & & & & D2 & $0.841$ ($0.104$) && $-0.660$ & $0.438$ \\
      & & \multirow{-4}{*}{$(500,100)$} & \multirow{-2}{*}{CAP-Med} & D4 & $0.898$ ($0.115$) && $-0.632$ & $0.402$ \\
      \cline{3-9}
      & & & & D2 & $0.915$ ($0.095$) && $-0.289$ & $0.087$ \\
      & & & \multirow{-2}{*}{GMed} & D4 & $0.985$ ($0.073$) && $-0.277$ & $0.080$ \\
      \cline{4-9}
      & & & & D2 & $0.905$ ($0.109$) && $-0.253$ & $0.071$ \\
      \multirow{-16}{*}{(1)} & \multirow{-8}{*}{$50$} & \multirow{-4}{*}{$(500,500)$} & \multirow{-2}{*}{CAP-Med} & D4 & $0.904$ ($0.137$) && $-0.129$ & $0.041$ \\
      \hline
      & & & & D2 & $0.993$ ($0.005$) && $-0.669$ & $0.449$ \\
      & & & \multirow{-2}{*}{GMed} & D4 & $0.999$ ($0.001$) && $-0.674$ & $0.456$ \\
      \cline{4-9}
      & & & & D2 & $0.993$ ($0.005$) && $-1.319$ & $1.863$ \\
      & & \multirow{-4}{*}{$(500,100)$} & \multirow{-2}{*}{GMed-Mis} & D4 & $0.999$ ($0.001$) && $-0.114$ & $0.014$ \\
      \cline{3-9}
      & & & & D2 & $0.999$ ($0.001$) && $-0.291$ & $0.087$ \\
      & & & \multirow{-2}{*}{GMed} & D4 & $1.000$ ($0.000$) && $-0.279$ & $0.081$ \\
      \cline{4-9}
      & & & & D2 & $0.999$ ($0.001$) && $-1.370$ & $2.076$ \\
      \multirow{-8}{*}{(2)} & \multirow{-8}{*}{$10$} & \multirow{-4}{*}{$(500,500)$} & \multirow{-2}{*}{GMed-Mis} & D4 & $1.000$ ($0.000$) && $-0.002$ & $0.001$ \\
      \hline
    \end{tabular}
  \end{center}
\end{table}

\section{Application}
\label{sec:application}

We apply the proposed approach to data acquired from the National Consortium on Alcohol and NeuroDevelopment in Adolescence (NCANDA). The consortium aims to identify the deterministic effect of alcohol use on the developing adolescent brain. A core battery of measurements, including structural and functional brain scans and cognitive testing, has been developed. In adolescence, the sex difference in cognitive behaviors has been identified in different functional domains, including spatial, verbal, math abilities, social recognition and so on~\citep{lauer2019development,meinhardt2020theory,esnaola2020development}. This difference has been shown to be partially mediated by brain functional connectivity captured by resting-state fMRI~\citep{alarcon2018adolescent}. In this study, we aim to identify the brain subnetwork within which functional connectivity mediates the sex difference in cognition. To exclude the impact of alcohol consumption, $n=621$ subjects ($312$ Males and $309$ Females) aged between $12$ and $22$ without excessive drinking are analyzed. Among these subjects, a significantly lower median response time in the motor speed test is observed in males compared to females after adjusting for age ($\mathrm{ATE}=-0.324$, $p\text{-value}<0.001$), where the motor task measures sensorimotor ability via having the participant use the mouse to click on a shrinking box when it moves to a new position on the screen. 
We further apply the proposed approach for mediation analysis, where sex is the binary exposure ($X$, $X=1$ for male), resting-state fMRI data are the mediator ($\bM$), the $z$-score of median response time for correct responses of the motor speed task is the outcome ($Y$), and age is the confounding factor ($W$). After preprocessing, fMRI time courses are extracted from $p=75$ brain regions ($60$ cortical and $15$ subcortical regions) spanning the whole brain using the Harvard-Oxford Atlas in FSL~\citep{smith2004advances}. These regions are grouped into $10$ functional modules, which will be used for an \textit{ad hoc} procedure of sparsifying the loading profile using the fused lasso~\citep{tibshirani2005sparsity} to impose local smoothness and constancy within each module. To remove the temporal dependence in the time courses, a subsample is taken with an effective sample size of $T_{i}=T=125$ and denote the subsampled data as $\bM_{i}\in\mathbb{R}^{T_{i}\times p}$, for $i=1,\dots,n$.

\begin{figure}
  \begin{center}
    \subfloat[Sparsified loading profile of $\btheta_{3}$]{\includegraphics[width=0.6\textwidth]{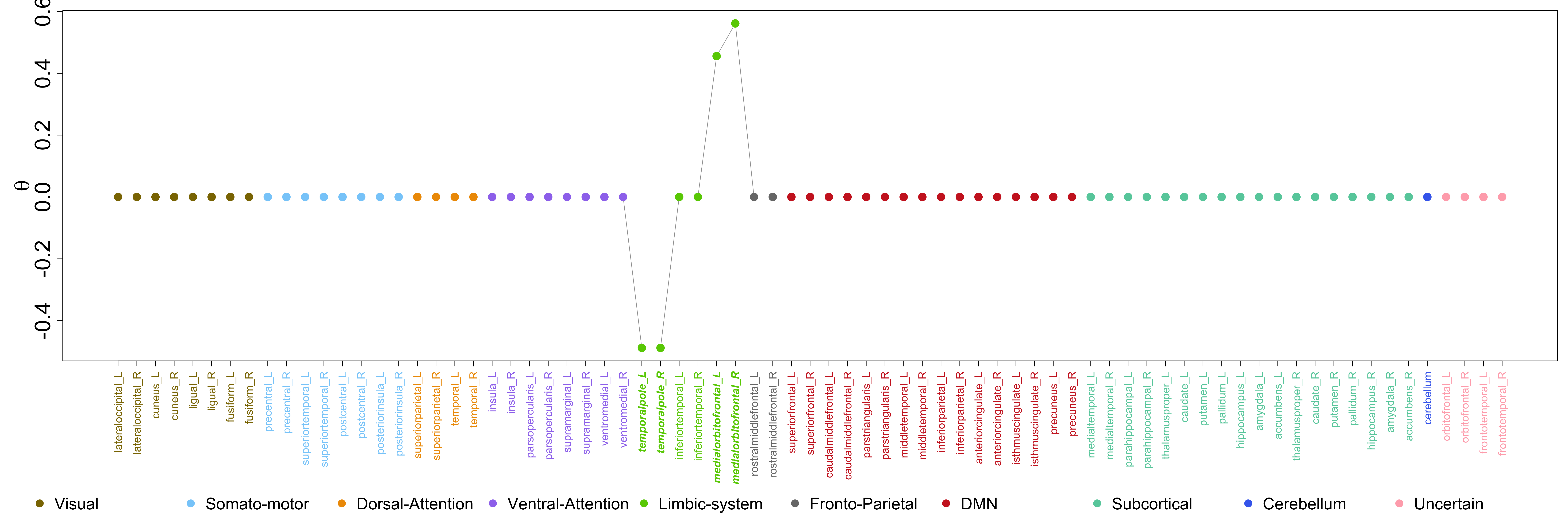}}
    \subfloat[Brain map of $\btheta_{3}$]{\includegraphics[width=0.3\textwidth]{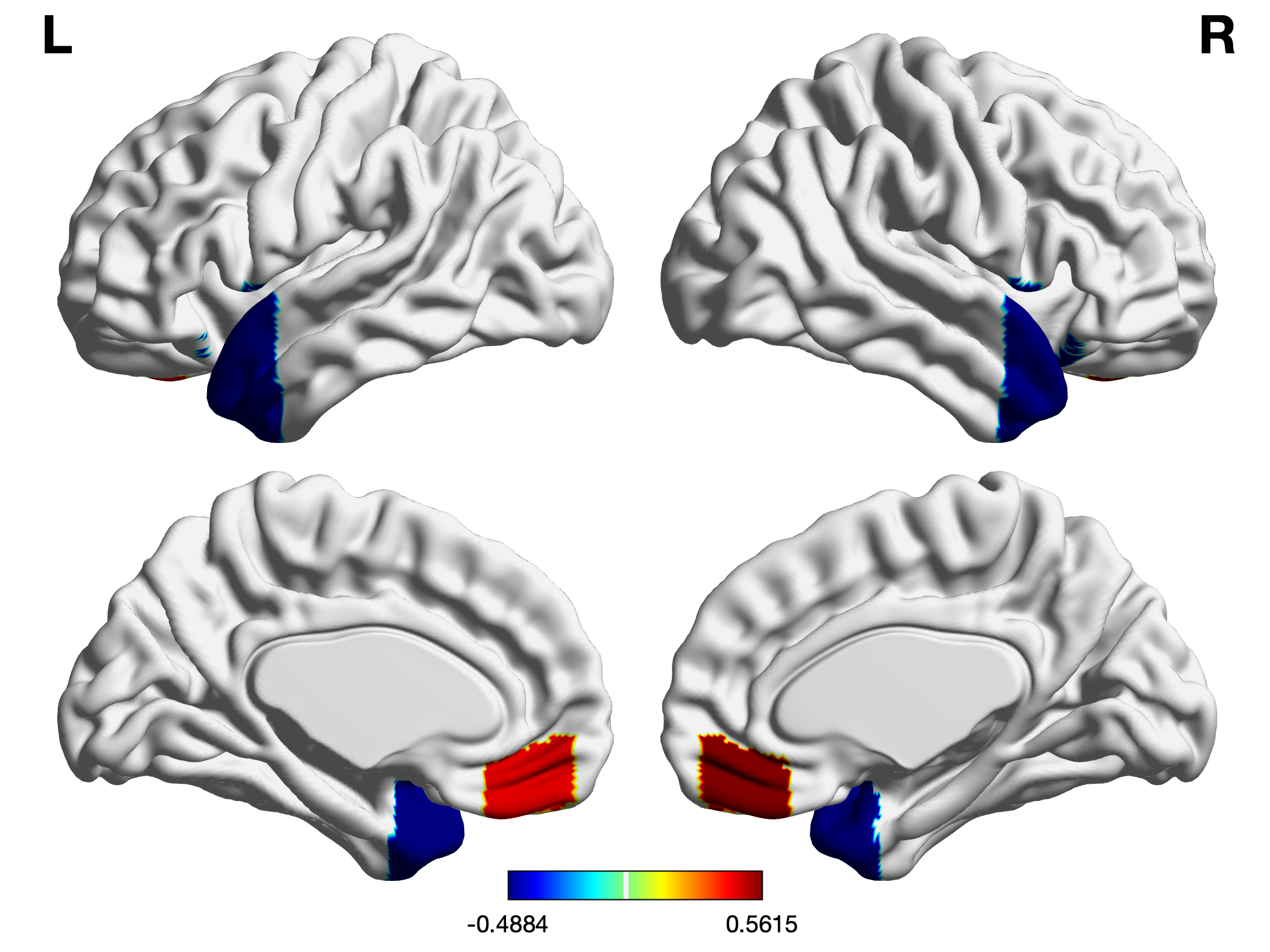}}
  \end{center}
  \caption{\label{fig:real_theta}The sparsified loading profile and brain map of the component with a significant AIE ($M_{3}$).}
\end{figure}

Using the $\mathrm{DfD}\leq 2$ criterion in~\eqref{eq:DfD}, the proposed approach identifies four components. Table~\ref{tabel:real_result} presents the estimated AIE, as well as $\alpha$ and $\beta$ coefficient. The confidence intervals are obtained from $500$ bootstrap samples. Among these identified components, the third component ($M_{3}$) shows a significantly positive AIE with both $\alpha$ and $\beta$ negative. Figure~\ref{fig:real_theta} presents the sparsified loading profile of $\btheta_{3}$ and the corresponding brain map. 
Section~\ref{appendix:sub:real_CAPMediation} of the supplementary materials presents the results from the CAP-Med approach introduced in Section~\ref{sec:sim}. The identified component with a significant mediation effect is consistent with $M_{3}$ identified by the proposed approach.
In $M_{3}$, four regions with a non-zero loading are all in the limbic-system network, including the temporal pole (left and right) and the medial orbitofrontal cortex (left and right). Compared to females, (weighted) functional connectivity within this network is lower in males, while this lower functional connectivity increases the reaction time. The temporal pole has been found associated with high-level cognitive functions~\citep{herlin2021temporal}. Though no direct relation to reaction time has been established, an indirect influence was hypothesized by contributing to processes involving decision-making, response selection, and emotion evaluation~\citep{pessoa2010emotion}. One of the primary functions of the medial orbitofrontal cortex is to integrate emotional reaction with sensory and/or contextual stimuli playing a role in reward processing and value-based decision making, which allows individuals to make adaptive responses to stimuli based on emotional significance~\citep{rudebeck2014orbitofrontal}. Thus, indirectly, activation in the area may prolong the reaction time. Regional sex differences of the temporal and frontal cortices have been observed in the developing brain using multiple imaging modalities. Sex difference in the brain was also suggested to be relevant to the symptomatic sex difference in psychiatric disorders~\citep{kaczkurkin2019sex}. Via a mediation analysis, the proposed approach offers a way of articulating the underlying mechanism.

\begin{table}
  \caption{\label{tabel:real_result}Estimated average indirect effect (AIE) and $\alpha$ and $\beta$ coefficient of the identified mediator components in the NCANDA dataset. Confidence intervals are constructed from $500$ bootstrap samples. Est.: estimate; SE: standard error; CI: confidence interval.}
  \begin{center}
    \resizebox{\textwidth}{!}{
    \begin{tabular}{l r c c c r c c c r c c}
      \hline
      & \multicolumn{3}{c}{AIE} && \multicolumn{3}{c}{$\alpha$} && \multicolumn{3}{c}{$\beta$} \\
      \cline{2-4}\cline{6-8}\cline{10-12}
      & \multicolumn{1}{c}{Est. (SE)} & \multicolumn{1}{c}{$p$-value} & \multicolumn{1}{c}{$95\%$ CI} && \multicolumn{1}{c}{Est. (SE)} & \multicolumn{1}{c}{$p$-value} & \multicolumn{1}{c}{$95\%$ CI} && \multicolumn{1}{c}{Est. (SE)} & \multicolumn{1}{c}{$p$-value} & \multicolumn{1}{c}{$95\%$ CI} \\
      \hline
      $M_{1}$ & $-0.018$ ($0.031$) & $0.574$ & $(-0.079, 0.044)$ && $-0.325$ ($0.069$) & $<0.001$ & $(-0.460, -0.190)$ && $0.053$ ($0.094$) & $0.572$ & $(-0.131, 0.237)$ \\
      $M_{2}$ & $-0.014$ ($0.032$) & $0.676$ & $(-0.077, 0.050)$ && $-0.286$ ($0.065$) & $<0.001$ & $(-0.414, -0.157)$ && $0.051$ ($0.110$) & $0.646$ & $(-0.166, 0.267)$ \\
      $M_{3}$ & $0.066$ ($0.027$) & $0.014$ & $(0.013, 0.119)$ && $-0.211$ ($0.064$) & $<0.001$ & $(-0.336, -0.086)$ && $-0.318$ ($0.094$) & $0.001$ & $(-0.503, -0.134)$ \\
      $M_{4}$ & $-0.035$ ($0.033$) & $0.287$ & $(-0.100, 0.030)$ && $0.256$ ($0.042$) & $<0.001$ & $(0.173, 0.338)$ && $-0.138$ ($0.127$) & $0.279$ & $(-0.388, 0.112)$ \\
      \hline
    \end{tabular}
    }
  \end{center}
\end{table}

\section{Discussion}
\label{sec:discussion}

This study introduces a mediation analysis framework when the mediator is a graph. A Gaussian covariance graph model is assumed for graph representation. Causal estimands and assumptions are discussed under this representation. With a covariance matrix as the mediator, parametric mediation models are considered based on matrix decomposition. Assuming Gaussian random errors, likelihood-based estimators are introduced to simultaneously identify the decomposition and causal parameters. An efficient computational algorithm is proposed and the asymptotic properties of the estimators are investigated. Via simulation studies, the performance of the proposed approach is evaluated. Applying to a resting-state fMRI study, a brain network is identified within which functional connectivity mediates the sex difference in the performance of a motor task.

In causal mediation analysis, an essential while untestable assumption is the assumption of no unmeasured mediator-outcome confounding. A sensitivity analysis is usually conducted to justify the validity of the conclusion to this assumption. For parametric approaches, one type of commonly used approach is to parametrize the confounding effect to evaluate the causal effects under various values, such as the one proposed in \citet{imai2010identification}. Using simulation studies, it is demonstrated that the proposed approach is robust to the existence of unmeasured mediator-outcome confounding in identifying the mediation component, that is in estimating the projection vector $\btheta$. With given $\btheta$, one can employ the approach in \citet{imai2010identification} for sensitivity analysis.

The asymptotic consistency of the proposed estimator requires the common diagonalization assumption on the covariance matrices. Via simulation studies, \citet{zhao2021covariate} pointed out that this assumption can be relaxed to partial common diagonalization. As this study also introduces a likelihood-based procedure, it is expected that the proposed approach is robust to this relaxation to partial common diagonalization.

Considering a graph mediator, under the Gaussian covariance graph model, this study assumes the number of nodes in the graph is fixed and low dimensional. The sample covariance matrices are thus well-conditioned and a likelihood-based approach is introduced to estimate model parameters. In many practical settings, for example, in voxel-level fMRI analysis, data dimension can be even higher than the number of fMRI data points. A well-conditioned estimator of the covariance matrix is required and we leave the introduction of such an estimator and the study of theoretical results as one of future directions.
As discussed in Section~\ref{sub:inference}, inference on projection vectors is not straightforward and requires rigorous theoretical and numerical investigations, which we leave to future research.




\appendix
\counterwithin{figure}{section}
\counterwithin{table}{section}
\counterwithin{equation}{section}
\counterwithin{lemma}{section}
\counterwithin{theorem}{section}.

\section*{Appendix}

This Appendix collects the technical proof of the theorems in the main text, additional technical details, and additional data analysis results.

\section{Theory and Proof}
\label{appendix:sec:proof}

\subsection{Proof of Theorem~\ref{thm:causal_estimand}}

\begin{proof}
  As discussed in Section~\ref{sub:causal_def}, the potential outcome of $Y$ under a multiple-worlds model is expressed as
  \[
    Y(x,\mathcal{G}(x'))=\gamma x+\alpha\beta x' +(\gamma_{0}+\alpha_{0}\beta+\beta\bW^\top\bphi_{1}+\bW^\top\bphi_{2})+\beta\eta+\epsilon.
  \]
  \begin{eqnarray*}
    \tau_{\mathrm{ATE}} &=& \mathbb{E}\left\{Y(1,\mathcal{G}(1))-Y(0,\mathcal{G}(0))\right\} \\
    &=& \left\{\gamma +\alpha\beta +(\gamma_{0}+\alpha_{0}\beta+\beta\bW^\top\bphi_{1}+\bW^\top\bphi_{2})\right\}-\left\{(\gamma_{0}+\alpha_{0}\beta+\beta\bW^\top\bphi_{1}+\bW^\top\bphi_{2})\right\} \\
    &=& \gamma+\alpha\beta.
  \end{eqnarray*}
  \begin{eqnarray*}
    \tau_{\mathrm{AIE}}(x) &=& \mathbb{E}\left\{Y(x,\mathcal{G}(1))-Y(x,\mathcal{G}(0))\right\} \\
    &=& \left\{\gamma x+\alpha\beta +(\gamma_{0}+\alpha_{0}\beta+\beta\bW^\top\bphi_{1}+\bW^\top\bphi_{2})\right\}-\left\{\gamma x +(\gamma_{0}+\alpha_{0}\beta+\beta\bW^\top\bphi_{1}+\bW^\top\bphi_{2})\right\} \\
    &=& \alpha\beta.
  \end{eqnarray*}
  \begin{eqnarray*}
    \tau_{\mathrm{ADE}}(x) &=& \mathbb{E}\left\{Y(1,\mathcal{G}(x))-Y(0,\mathcal{G}(x))\right\} \\
    &=& \left\{\gamma+\alpha\beta x +(\gamma_{0}+\alpha_{0}\beta+\beta\bW^\top\bphi_{1}+\bW^\top\bphi_{2})\right\}-\left\{\alpha\beta x +(\gamma_{0}+\alpha_{0}\beta+\beta\bW^\top\bphi_{1}+\bW^\top\bphi_{2})\right\} \\
    &=& \gamma.
  \end{eqnarray*}
  The above proves Theorem~\ref{thm:causal_estimand}.
\end{proof}

\subsection{Proof of Lemma~\ref{lemma:hlikelihood_const}}

From the lemma assumption, $\hat{\bSigma}_{i}$ is a consistent estimator of $\bSigma_{i}$. With given $\btheta$, the likelihood function is a continuous function of $\hat{\bSigma}_{i}$ (for $i=1,\dots,n$). Thus, $\hat{\ell}$ converges to $\ell$ as $\min_{i}T_{i}\rightarrow\infty$.

\subsection{Proof of Proposition~\ref{prop:est_asmp}}

Assumption B1 assumes a low-dimensional scenario and the dimension of the mediator is fixed. Lemma~\ref{lemma:hlikelihood_const} demonstrates the consistency of the approximated likelihood function used for optimization. Asymptotically, the proposed estimator is likelihood-based. Thus, under the imposed assumptions, the consistency of the estimator follows.

\section{Details of Algorithm~\ref{alg:capmed}}
\label{appendix:sec:alg_covreg}

\begin{eqnarray*}
  \hat{\ell} &=& \sum_{i=1}^{n}\frac{T_{i}}{2}\left\{(\alpha_{0i}+\bX_{i}^\top\balpha)+(\btheta^\top\bS_{i}\btheta)\exp(-\alpha_{0i}-\bX_{i}^\top\balpha)\right\} \\
  && +\sum_{i=1}^{n}\frac{1}{2}\left\{\log\sigma^{2}+\frac{1}{\sigma^{2}}(Y_{i}-\gamma_{0}-\bX_{i}^\top\bgamma-\beta\log(\btheta^\top\hat{\bSigma}_{i}\btheta))^{2}\right\}+\sum_{i=1}^{n}\frac{1}{2}\left\{\log\pi^{2}+\frac{1}{\pi^{2}}(\alpha_{0i}-\alpha_{0})^{2}\right\}. \nonumber
\end{eqnarray*}

First considering $\balpha$ and $\alpha_{0i}$ ($i=1,\dots,n$), analytical solution is not available, thus the Newton-Raphson algorithm is employed. 
For $\balpha$,
 \[
  \frac{\partial\hat{\ell}}{\partial\balpha}=\frac{1}{2}\sum_{i=1}^{n}T_{i}\left\{1-(\btheta^\top\bS_{i}\btheta)\exp(-\alpha_{0i}-\bX_{i}^\top\balpha)\right\}\bX_{i},
 \]
 \[
  \frac{\partial^{2}\hat{\ell}}{\partial\balpha\partial\balpha^\top}=\frac{1}{2}\sum_{i=1}^{n}T_{i}(\btheta^\top\bS_{i}\btheta)\exp(-\alpha_{0i}-\bX_{i}^\top\balpha)\bX_{i}\bX_{i}^\top,
 \]
 \[
  \balpha^{(s+1)}=\balpha^{(s)}-\left(\left. \frac{\partial^{2}\hat{\ell}}{\partial\balpha\partial\balpha^\top} \right\vert_{\balpha^{(s)}}\right)^{-1}\left(\left. \frac{\partial\hat{\ell}}{\partial\balpha} \right\vert_{\balpha^{(s)}}\right).
 \]
For $\alpha_{0i}$ ($i=1,\dots,n$),
 \[
  \frac{\partial\hat{\ell}}{\partial\alpha_{0i}}=\frac{1}{2}\left[T_{i}\left\{1-(\btheta^\top\bS_{i}\btheta)\exp(-\alpha_{0i}-\bX_{i}^\top\balpha)\right\}+\frac{2}{\tau^{2}}(\alpha_{0i}-\alpha_{0}) \right],
 \]
 \[
  \frac{\partial^{2}\hat{\ell}}{\partial\alpha_{0i}^{2}}=\frac{1}{2}\left\{T_{i}(\btheta^\top\bS_{i}\btheta)\exp(-\alpha_{0i}-\bX_{i}^\top\balpha)+\frac{2}{\tau^{2}}\right\},
 \]
 \[
  \alpha_{0i}^{(s+1)}=\alpha_{0i}^{(s)}-\left(\left. \frac{\partial^{2}\hat{\ell}}{\partial\alpha_{0i}^{2}} \right\vert_{\alpha_{0i}^{(s)}}\right)^{-1}\left(\left. \frac{\partial\hat{\ell}}{\partial\alpha_{0i}} \right\vert_{\alpha_{0i}^{(s)}}\right)
 \]
For $\alpha_{0}$ and $\tau$, explicit form for update is available,
 \[
  \alpha_{0}^{(s+1)}=\frac{1}{n}\sum_{i=1}^{n}\alpha_{0i}^{(s+1)}, \quad \tau^{2(s+1)}=\frac{1}{n}\sum_{i=1}^{n}\left(\alpha_{0i}^{(s+1)}-\alpha_{0}^{(s+1)}\right)^{2}.
 \]
For $(\gamma_{0},\bgamma,\beta)$, the analytical solution for update can be found jointly. Let
 \[
  \bZ_{i}=\begin{pmatrix}
   1 \\
   \bX_{i} \\
   \log(\btheta^\top\hat{\bSigma}_{i}\btheta)
  \end{pmatrix}, \quad \bmu=\begin{pmatrix}
   \gamma_{0} \\
   \bgamma \\
   \beta
  \end{pmatrix},
 \]
 \[
  \frac{\partial\hat{\ell}}{\partial\bmu}=\frac{1}{2}\sum_{i=1}^{n}\frac{2}{\sigma^{2}}(Y_{i}-\bZ_{i}^\top\bmu)(-\bZ_{i})=\boldsymbol{\mathrm{0}},
 \]
 \[
  \Rightarrow \quad \bmu^{(s+1)}=\left(\sum_{i=1}^{n}\bZ_{i}^{(s)}\bZ_{i}^{(s)\top}\right)^{-1}\left(\sum_{i=1}^{n}Y_{i}\bZ_{i}^{(s)}\right)=(\bZ^{(s)\top}\bZ^{(s)})^{-1}\bZ^{(s)\top}\bY,
 \]
where $\bZ=(\bZ_{1},\dots,\bZ_{n})^\top$.
For $\sigma$,
 \[
  \sigma^{2(s+1)}=\frac{1}{n}\sum_{i=1}^{n}(Y_{i}-\bZ_{i}^{(s)\top}\bmu^{(s+1)})^{2}.
 \]

For $\btheta$, to find the solution to~\eqref{eq:solve_theta}, it is equivalent to find the eigenvectors and eigenvalues of $\bA$ with respect to $\bH$. We first assume $\btheta_{0}$ is a solution eigenvector with unit norm, $\|\btheta_{0}\|_{2}=1$. Since $\bH$ is positive definite, let $\btheta=\bH^{-1 / 2} \btheta_{0}$, then
\[
  \btheta^{\top} \bH \btheta=\btheta_{0}^{\top} \bH^{-1 / 2} \bH \bH^{-1 / 2} \btheta_{0}=\btheta_{0}^{\top} \btheta_{0}=1,
\]
which satisfies the constraint condition. Replace $\btheta$ with $\btheta=\bH^{-1 / 2} \btheta_{0}$ in~\eqref{eq:solve_theta},
\[
\bA\bH^{-1 / 2} \btheta_{0}-\lambda \bH\bH^{-1 / 2} \btheta_0=\boldsymbol{\mathrm{0}}.
\]
\[
\Rightarrow \quad \bH^{-1 / 2} \bA\bH^{-1 / 2} \btheta_{0}=\lambda \btheta_{0}.
\]
Therefore, $\btheta_{0}$ is the eigenvector of matrix $\bH^{-1 / 2} \bA \bH^{-1 / 2}$ and $\lambda$ is the corresponding eigenvalue. Then, $\btheta=\bH^{-1/2}\btheta_{0}$ is the update for $\btheta$.


\section{Additional Results of the NCANDA Study}
\label{appendix:sec:real}

\subsection{Sensitivity Analysis}
\label{appendix:sub:real_sens}

\begin{figure}
  \begin{center}
    \includegraphics[width=0.5\textwidth]{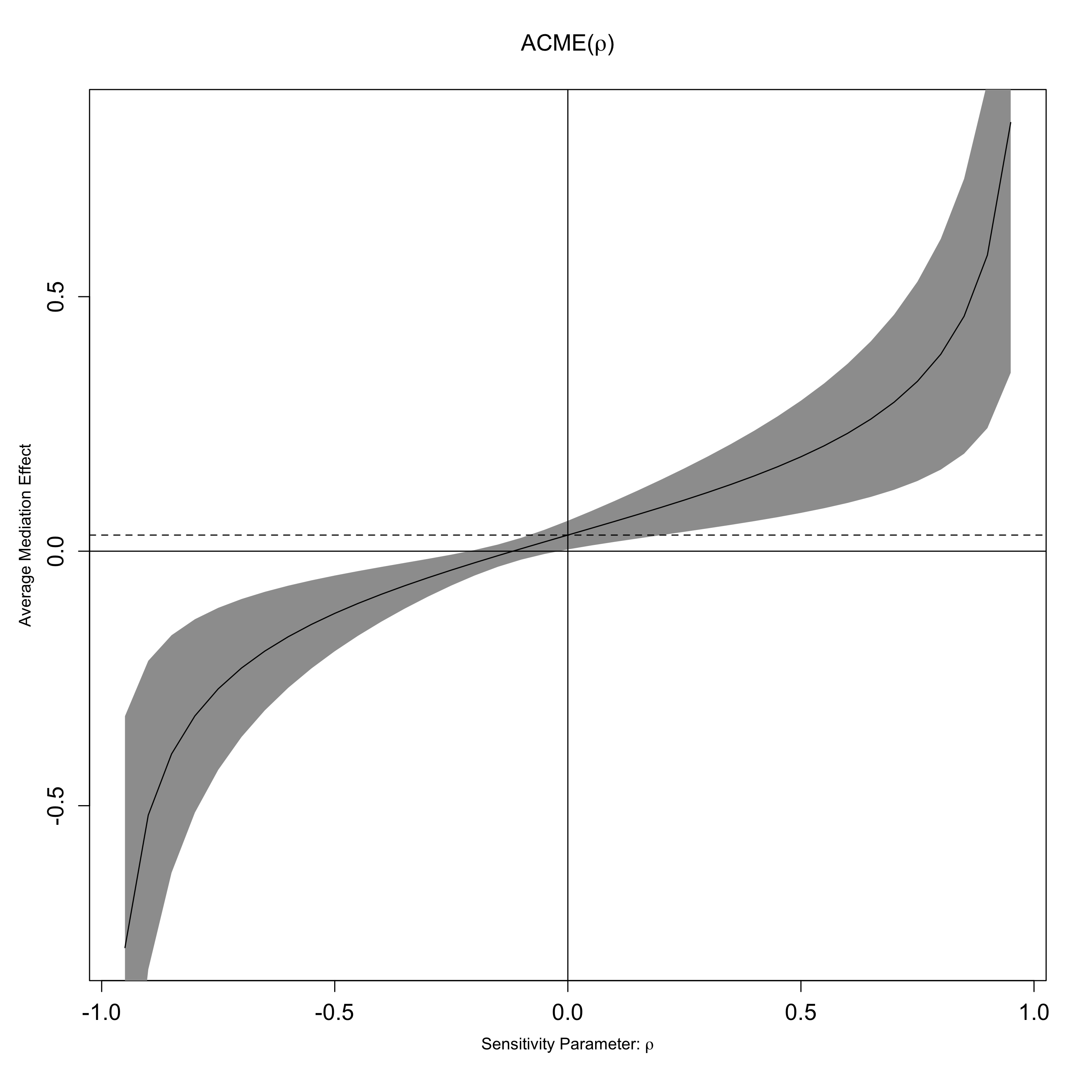}
  \end{center}
  \caption{\label{fig:real_M3_sens}Sensitivity analysis of $M_{3}$ in the NCANDA study.}
\end{figure}

As discussed in Section~\ref{sec:sim}, the proposed approach is robust to the existence of an additive unmeasured mediator-outcome confounding in identifying the projection vectors. Based on this property, a sensitivity analysis introduced in \citet{imai2010identification} is implemented where for an additive unmeasured mediator-outcome confounding, the correlation between the model errors is considered as the sensitivity parameter. In the NCANDA application study, $M_{3}$ shows a significant mediation effect. With the estimated $\btheta$ of $M_{3}$, Figure~\ref{fig:real_M3_sens} presents the sensitivity analysis plot over the range of the sensitivity parameter ($\rho$). In the figure, the $95\%$ confidence intervals are constructed from $500$ bootstrap samples. From the figure, the $95\%$ confidence interval covers zero for the $\rho$ value between $-0.20$ and $-0.05$. When $\rho>-0.05$, the average indirect effect is positive and significant; when $\rho < -0.20$, the average indirect effect is negative and significant.

\subsection{Results from the CAP-Med approach}
\label{appendix:sub:real_CAPMediation}

\begin{figure}
  \begin{center}
    \subfloat[Sparsified loading profile of C7]{\includegraphics[width=0.6\textwidth]{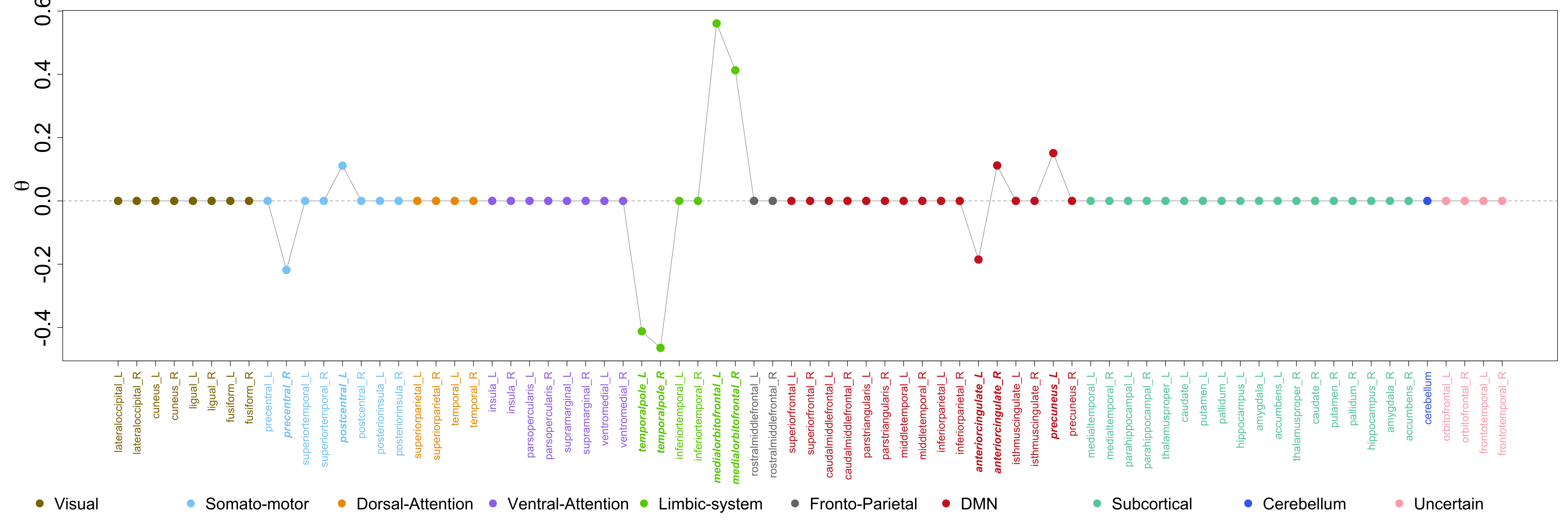}}
    \subfloat[Brain map of C7]{\includegraphics[width=0.3\textwidth]{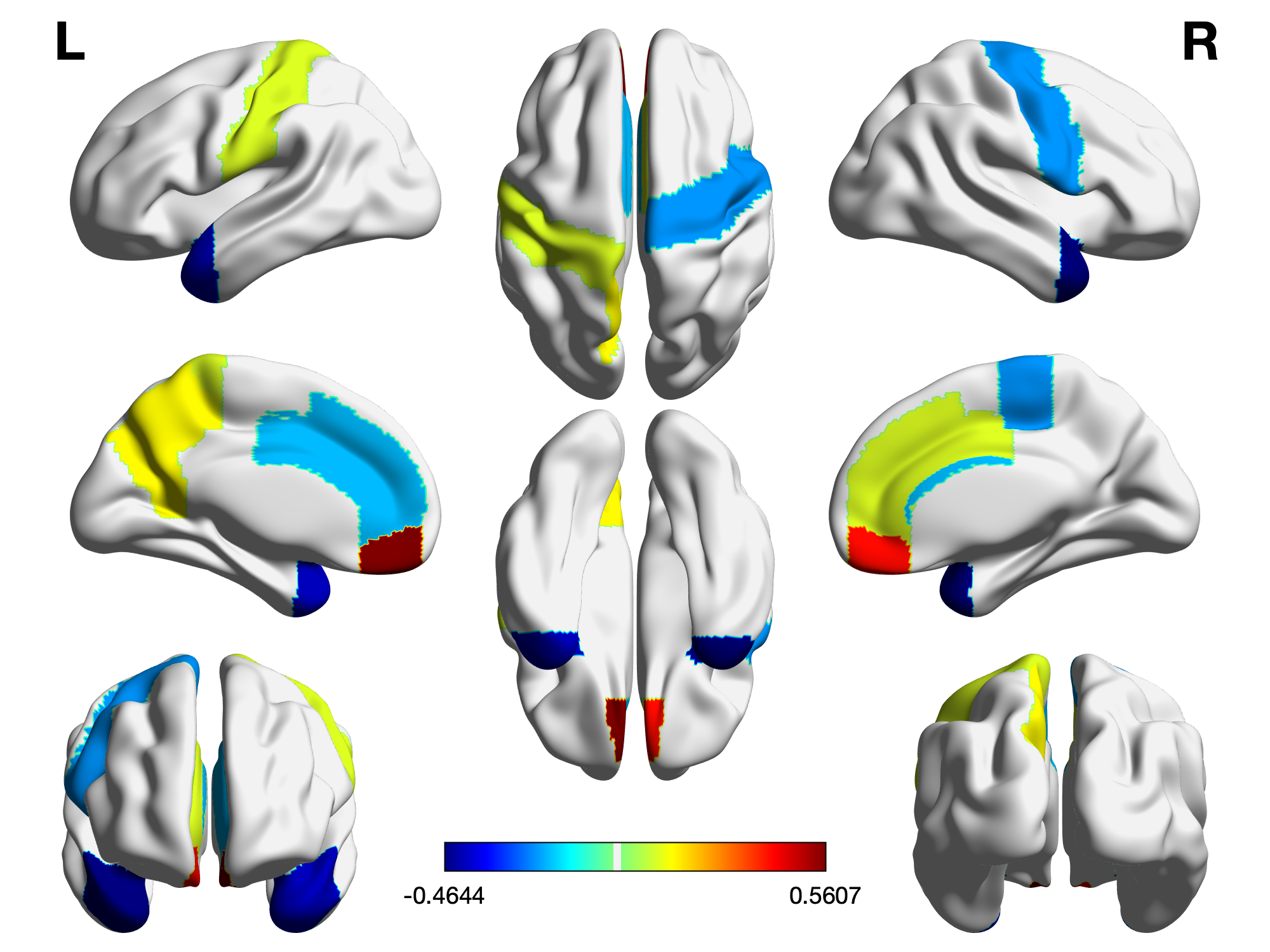}}
  \end{center}
  \caption{\label{fig:real_CAPMediation_C7}The sparsified loading profile and brain map of the component with a significant AIE (C7) using the CAP-Med approach.}
\end{figure}

We also apply the CAP-Med approach introduced in Section~\ref{sec:sim} to the NCANDA data. The CAP step identifies $8$ components using the criterion of $\mathrm{DfD}\leq 2$. After running the mediation analysis, the seventh component (C7) shows a significantly positive average indirect effect with $\mathrm{AIE}=0.038$ and $95\%$ CI $(0.004,0.078)$.
Figure~\ref{fig:real_CAPMediation_C7} presents the sparsified loading profile of this component and the corresponding regions in the brain map. Compared to the profile of $M_{3}$ identified by the proposed approach, GMed, a high similarity between the two is observed with a similarity of $|\langle \hat{\btheta}_{3}^{\text{(GMed)}},\hat{\btheta}_{7}^{\text{(CAP-Med)}}\rangle|=0.864$. For the rest components identified by GMed, a corresponding component from CAP-Med with high similarity can be found: $|\langle \hat{\btheta}_{1}^{\text{(GMed)}},\hat{\btheta}_{8}^{\text{(CAP-Med)}}\rangle|=0.805$, $|\langle \hat{\btheta}_{2}^{\text{(GMed)}},\hat{\btheta}_{6}^{\text{(CAP-Med)}}\rangle|=0.772$, and $|\langle \hat{\btheta}_{4}^{\text{(GMed)}},\hat{\btheta}_{2}^{\text{(CAP-Med)}}\rangle|=0.862$.
Though both approaches identify these consistent components, the proposed approach offers an integrated way of targeting the ones demonstrating a mediation effect rather than performing a two-step approach.

\bibliographystyle{apalike}
\bibliography{Bibliography}

\begin{thebibliography}{}

\bibitem[Alarc{\'o}n et~al., 2018]{alarcon2018adolescent}
Alarc{\'o}n, G., Pfeifer, J.~H., Fair, D.~A., and Nagel, B.~J. (2018).
\newblock Adolescent gender differences in cognitive control performance and
  functional connectivity between default mode and fronto-parietal networks
  within a self-referential context.
\newblock {\em Frontiers in Behavioral Neuroscience}, 12:73.

\bibitem[Anderson, 1973]{anderson1973asymptotically}
Anderson, T. (1973).
\newblock Asymptotically efficient estimation of covariance matrices with
  linear structure.
\newblock {\em The Annals of Statistics}, 1(1):135--141.

\bibitem[Andrews and Didelez, 2020]{andrews2020insights}
Andrews, R.~M. and Didelez, V. (2020).
\newblock Insights into the cross-world independence assumption of causal
  mediation analysis.
\newblock {\em Epidemiology}, 32(2):209--219.

\bibitem[Baron and Kenny, 1986]{baron1986moderator}
Baron, R.~M. and Kenny, D.~A. (1986).
\newblock The moderator--mediator variable distinction in social psychological
  research: Conceptual, strategic, and statistical considerations.
\newblock {\em Journal of Personality and Social Psychology}, 51(6):1173.

\bibitem[Boik, 2002]{boik2002spectral}
Boik, R.~J. (2002).
\newblock Spectral models for covariance matrices.
\newblock {\em Biometrika}, 89(1):159--182.

\bibitem[Chaudhuri et~al., 2007]{chaudhuri2007estimation}
Chaudhuri, S., Drton, M., and Richardson, T.~S. (2007).
\newblock Estimation of a covariance matrix with zeros.
\newblock {\em Biometrika}, 94(1):199--216.

\bibitem[Chen and Zhou, 2023]{chen2023causal}
Chen, M. and Zhou, Y. (2023).
\newblock Causal mediation analysis with a three-dimensional image mediator.
\newblock {\em arXiv preprint arXiv:2303.06560}.

\bibitem[Ch{\'e}n et~al., 2017]{chen2017high}
Ch{\'e}n, O.~Y., Crainiceanu, C., Ogburn, E.~L., Caffo, B.~S., Wager, T.~D.,
  and Lindquist, M.~A. (2017).
\newblock High-dimensional multivariate mediation with application to
  neuroimaging data.
\newblock {\em Biostatistics}, 19(2):121--136.

\bibitem[Chiu et~al., 1996]{chiu1996matrix}
Chiu, T.~Y., Leonard, T., and Tsui, K.-W. (1996).
\newblock The matrix-logarithmic covariance model.
\newblock {\em Journal of the American Statistical Association},
  91(433):198--210.

\bibitem[Derkach et~al., 2019]{derkach2019high}
Derkach, A., Pfeiffer, R.~M., Chen, T.-H., and Sampson, J.~N. (2019).
\newblock High dimensional mediation analysis with latent variables.
\newblock {\em Biometrics}, 75(3):745--756.

\bibitem[Edwards, 2012]{edwards2012introduction}
Edwards, D. (2012).
\newblock {\em Introduction to graphical modelling}.
\newblock Springer Science \& Business Media.

\bibitem[Esnaola et~al., 2020]{esnaola2020development}
Esnaola, I., Ses{\'e}, A., Antonio-Agirre, I., and Azpiazu, L. (2020).
\newblock The development of multiple self-concept dimensions during
  adolescence.
\newblock {\em Journal of Research on Adolescence}, 30:100--114.

\bibitem[Flury, 1984]{flury1984common}
Flury, B.~N. (1984).
\newblock Common principal components in $k$ groups.
\newblock {\em Journal of the American Statistical Association},
  79(388):892--898.

\bibitem[Franks and Hoff, 2019]{franks2019shared}
Franks, A.~M. and Hoff, P. (2019).
\newblock Shared subspace models for multi-group covariance estimation.
\newblock {\em Journal of Machine Learning Research}, 20(171):1--37.

\bibitem[Friston, 2011]{friston2011functional}
Friston, K.~J. (2011).
\newblock Functional and effective connectivity: a review.
\newblock {\em Brain Connectivity}, 1(1):13--36.

\bibitem[Gu et~al., 2014]{gu2014state}
Gu, F., Preacher, K.~J., and Ferrer, E. (2014).
\newblock A state space modeling approach to mediation analysis.
\newblock {\em Journal of Educational and Behavioral Statistics},
  39(2):117--143.

\bibitem[Herlin et~al., 2021]{herlin2021temporal}
Herlin, B., Navarro, V., and Dupont, S. (2021).
\newblock The temporal pole: From anatomy to function—a literature appraisal.
\newblock {\em Journal of Chemical Neuroanatomy}, 113:101925.

\bibitem[Hoff, 2009]{hoff2009hierarchical}
Hoff, P.~D. (2009).
\newblock A hierarchical eigenmodel for pooled covariance estimation.
\newblock {\em Journal of the Royal Statistical Society: Series B (Statistical
  Methodology)}, 71(5):971--992.

\bibitem[Hoff and Niu, 2012]{hoff2012covariance}
Hoff, P.~D. and Niu, X. (2012).
\newblock A covariance regression model.
\newblock {\em Statistica Sinica}, 22(2):729--753.

\bibitem[Huang and Pan, 2016]{huang2016hypothesis}
Huang, Y.-T. and Pan, W.-C. (2016).
\newblock Hypothesis test of mediation effect in causal mediation model with
  high-dimensional continuous mediators.
\newblock {\em Biometrics}, 72(2):402--413.

\bibitem[Imai et~al., 2010]{imai2010identification}
Imai, K., Keele, L., and Yamamoto, T. (2010).
\newblock Identification, inference and sensitivity analysis for causal
  mediation effects.
\newblock {\em Statistical Science}, 25(1):51--71.

\bibitem[Jiang and Colditz, 2023]{jiang2023causal}
Jiang, S. and Colditz, G.~A. (2023).
\newblock Causal mediation analysis using high-dimensional image mediator
  bounded in irregular domain with an application to breast cancer.
\newblock {\em Biometrics}.

\bibitem[Kaczkurkin et~al., 2019]{kaczkurkin2019sex}
Kaczkurkin, A.~N., Raznahan, A., and Satterthwaite, T.~D. (2019).
\newblock Sex differences in the developing brain: insights from multimodal
  neuroimaging.
\newblock {\em Neuropsychopharmacology}, 44(1):71--85.

\bibitem[Krzanowski, 1984]{krzanowski1984principal}
Krzanowski, W. (1984).
\newblock Principal component analysis in the presence of group structure.
\newblock {\em Journal of the Royal Statistical Society: Series C (Applied
  Statistics)}, 33(2):164--168.

\bibitem[Lauer et~al., 2019]{lauer2019development}
Lauer, J.~E., Yhang, E., and Lourenco, S.~F. (2019).
\newblock The development of gender differences in spatial reasoning: A
  meta-analytic review.
\newblock {\em Psychological Bulletin}, 145(6):537.

\bibitem[Lee and Nelder, 1996]{lee1996hierarchical}
Lee, Y. and Nelder, J.~A. (1996).
\newblock Hierarchical generalized linear models.
\newblock {\em Journal of the Royal Statistical Society. Series B
  (Methodological)}, pages 619--678.

\bibitem[Lindquist, 2008]{lindquist2008statistical}
Lindquist, M.~A. (2008).
\newblock The statistical analysis of {fMRI} data.
\newblock {\em Statistical Science}, 23(4):439--464.

\bibitem[Lindquist, 2012]{lindquist2012functional}
Lindquist, M.~A. (2012).
\newblock Functional causal mediation analysis with an application to brain
  connectivity.
\newblock {\em Journal of the American Statistical Association},
  107(500):1297--1309.

\bibitem[Meinhardt-Injac et~al., 2020]{meinhardt2020theory}
Meinhardt-Injac, B., Daum, M.~M., and Meinhardt, G. (2020).
\newblock Theory of mind development from adolescence to adulthood: Testing the
  two-component model.
\newblock {\em British Journal of Developmental Psychology}, 38(2):289--303.

\bibitem[Pearl, 2001]{pearl2001direct}
Pearl, J. (2001).
\newblock Direct and indirect effects.
\newblock In {\em Proceedings of the seventeenth conference on uncertainty in
  artificial intelligence}, pages 411--420. Morgan Kaufmann Publishers Inc.

\bibitem[Pessoa, 2010]{pessoa2010emotion}
Pessoa, L. (2010).
\newblock Emotion and cognition and the amygdala: from ``what is it?'' to
  ``what's to be done?''.
\newblock {\em Neuropsychologia}, 48(12):3416--3429.

\bibitem[Pourahmadi et~al., 2007]{pourahmadi2007simultaneous}
Pourahmadi, M., Daniels, M.~J., and Park, T. (2007).
\newblock Simultaneous modelling of the {Cholesky} decomposition of several
  covariance matrices.
\newblock {\em Journal of Multivariate Analysis}, 98(3):568--587.

\bibitem[Richardson and Spirtes, 2002]{richardson2002ancestral}
Richardson, T. and Spirtes, P. (2002).
\newblock Ancestral graph markov models.
\newblock {\em Annals of Statistics}, pages 962--1030.

\bibitem[Richardson and Robins, 2013]{richardson2013single}
Richardson, T.~S. and Robins, J.~M. (2013).
\newblock Single world intervention graphs (swigs): A unification of the
  counterfactual and graphical approaches to causality.
\newblock {\em Center for the Statistics and the Social Sciences, University of
  Washington Series. Working Paper}, 128(30):2013.

\bibitem[Robins and Greenland, 1992]{robins1992identifiability}
Robins, J.~M. and Greenland, S. (1992).
\newblock Identifiability and exchangeability for direct and indirect effects.
\newblock {\em Epidemiology}, pages 143--155.

\bibitem[Robins and Richardson, 2010]{robins2010alternative}
Robins, J.~M. and Richardson, T.~S. (2010).
\newblock Alternative graphical causal models and the identification of direct
  effects.
\newblock {\em Causality and psychopathology: Finding the determinants of
  disorders and their cures}, pages 103--158.

\bibitem[Rubin, 1980]{rubin1980randomization}
Rubin, D.~B. (1980).
\newblock Randomization analysis of experimental data: The {Fisher}
  randomization test comment.
\newblock {\em Journal of the American Statistical Association},
  75(371):591--593.

\bibitem[Rudebeck and Murray, 2014]{rudebeck2014orbitofrontal}
Rudebeck, P.~H. and Murray, E.~A. (2014).
\newblock The orbitofrontal oracle: cortical mechanisms for the prediction and
  evaluation of specific behavioral outcomes.
\newblock {\em Neuron}, 84(6):1143--1156.

\bibitem[Seiler and Holmes, 2017]{seiler2017multivariate}
Seiler, C. and Holmes, S. (2017).
\newblock Multivariate heteroscedasticity models for functional brain
  connectivity.
\newblock {\em Frontiers in Neuroscience}, 11:696.

\bibitem[Smith et~al., 2004]{smith2004advances}
Smith, S.~M., Jenkinson, M., Woolrich, M.~W., Beckmann, C.~F., Behrens, T.~E.,
  Johansen-Berg, H., Bannister, P.~R., De~Luca, M., Drobnjak, I., Flitney,
  D.~E., et~al. (2004).
\newblock Advances in functional and structural {MR} image analysis and
  implementation as {FSL}.
\newblock {\em NeuroImage}, 23:S208--S219.

\bibitem[Tibshirani et~al., 2005]{tibshirani2005sparsity}
Tibshirani, R., Saunders, M., Rosset, S., Zhu, J., and Knight, K. (2005).
\newblock Sparsity and smoothness via the fused lasso.
\newblock {\em Journal of the Royal Statistical Society: Series B (Statistical
  Methodology)}, 67(1):91--108.

\bibitem[VanderWeele, 2015]{vanderweele2015explanation}
VanderWeele, T.~J. (2015).
\newblock {\em Explanation in Causal Inference: Methods for Mediation and
  Interaction}.
\newblock Oxford University Press.

\bibitem[VanderWeele, 2016]{vanderweele2016mediation}
VanderWeele, T.~J. (2016).
\newblock Mediation analysis: a practitioner's guide.
\newblock {\em Annual review of public health}, 37:17--32.

\bibitem[Zeng et~al., 2021]{zeng2021causal}
Zeng, S., Rosenbaum, S., Alberts, S.~C., Archie, E.~A., and Li, F. (2021).
\newblock Causal mediation analysis for sparse and irregular longitudinal data.
\newblock {\em The Annals of Applied Statistics}, 15(2):747--767.

\bibitem[Zhao et~al., 2022]{zhao2022bayesian}
Zhao, Y., Chen, T., Cai, J., Lichenstein, S., Potenza, M.~N., and Yip, S.~W.
  (2022).
\newblock {Bayesian} network mediation analysis with application to the brain
  functional connectome.
\newblock {\em Statistics in Medicine}, 41(20):3991--4005.

\bibitem[Zhao and Luo, 2019]{zhao2019granger}
Zhao, Y. and Luo, X. (2019).
\newblock Granger mediation analysis of multiple time series with an
  application to functional magnetic resonance imaging.
\newblock {\em Biometrics}, 75(3):788--798.

\bibitem[Zhao and Luo, 2022]{zhao2022pathway}
Zhao, Y. and Luo, X. (2022).
\newblock Pathway lasso: Pathway estimation and selection with high dimensional
  mediators.
\newblock {\em Statistics and Its Interface}, 15(1):39--50.

\bibitem[Zhao et~al., 2018]{zhao2018functional}
Zhao, Y., Luo, X., Lindquist, M., and Caffo, B. (2018).
\newblock Functional mediation analysis with an application to functional
  magnetic resonance imaging data.
\newblock {\em arXiv preprint arXiv:1805.06923}.

\bibitem[Zhao et~al., 2021]{zhao2021covariate}
Zhao, Y., Wang, B., Mostofsky, S.~H., Caffo, B.~S., and Luo, X. (2021).
\newblock Covariate assisted principal regression for covariance matrix
  outcomes.
\newblock {\em Biostatistics}, 22(3):629--645.

\bibitem[Zou et~al., 2017]{zou2017covariance}
Zou, T., Lan, W., Wang, H., and Tsai, C.-L. (2017).
\newblock Covariance regression analysis.
\newblock {\em Journal of the American Statistical Association},
  112(517):266--281.

\end{thebibliography}

\end{document}